\documentclass[10pt,twocolumn,twoside]{IEEEtran}

\usepackage{amssymb}
\usepackage{amsfonts}
\usepackage{amsmath}
\usepackage{cite}
\usepackage[dvips]{graphicx}
\usepackage{color}
\usepackage{algorithm}
\usepackage{algorithmic}
\usepackage{verbatim}

\begin{document}

\newtheorem{definition}{\it Definition}
\newtheorem{theorem}{\bf Theorem}
\newtheorem{lemma}{\it Lemma}
\newtheorem{corollary}{\it Corollary}
\newtheorem{remark}{\it Remark}
\newtheorem{example}{\it Example}
\newtheorem{case}{\bf Case Study}
\newtheorem{assumption}{\it Assumption}
\newtheorem{property}{\it Property}
\newtheorem{proposition}{\it Proposition}

\newcommand{\hP}[1]{{\boldsymbol h}_{{#1}{\bullet}}}
\newcommand{\hS}[1]{{\boldsymbol h}_{{\bullet}{#1}}}

\newcommand{\ba}{\boldsymbol{a}}
\newcommand{\baq}{\overline{q}}
\newcommand{\bA}{\boldsymbol{A}}
\newcommand{\bb}{\boldsymbol{b}}
\newcommand{\bB}{\boldsymbol{B}}
\newcommand{\bc}{\boldsymbol{c}}
\newcommand{\bcC}{\boldsymbol{\cal C}}
\newcommand{\bcO}{\boldsymbol{\cal O}}
\newcommand{\bd}{\boldsymbol{d}}
\newcommand{\bh}{\boldsymbol{h}}
\newcommand{\bH}{\boldsymbol{H}}
\newcommand{\bl}{\boldsymbol{l}}
\newcommand{\bm}{\boldsymbol{m}}
\newcommand{\bn}{\boldsymbol{n}}
\newcommand{\bo}{\boldsymbol{o}}
\newcommand{\bO}{\boldsymbol{O}}
\newcommand{\bp}{\boldsymbol{p}}
\newcommand{\bq}{\boldsymbol{q}}
\newcommand{\br}{\boldsymbol{r}}
\newcommand{\bR}{\boldsymbol{R}}
\newcommand{\bs}{\boldsymbol{s}}
\newcommand{\bS}{\boldsymbol{S}}
\newcommand{\bT}{\boldsymbol{T}}
\newcommand{\bu}{\boldsymbol{u}}
\newcommand{\bv}{\boldsymbol{v}}
\newcommand{\bw}{\boldsymbol{w}}
\newcommand{\bX}{\boldsymbol{X}}
\newcommand{\bZ}{\boldsymbol{Z}}
\newcommand{\bzero}{\boldsymbol{0}}

\newcommand{\balpha}{\boldsymbol{\alpha}}
\newcommand{\bbeta}{\boldsymbol{\beta}}
\newcommand{\bdeta}{\boldsymbol{\eta}}
\newcommand{\btau}{\boldsymbol{\tau}}
\newcommand{\bOmega}{\boldsymbol{\Omega}}
\newcommand{\bTheta}{\boldsymbol{\Theta}}
\newcommand{\bLambda}{\boldsymbol{\Lambda}}
\newcommand{\bphi}{\boldsymbol{\phi}}
\newcommand{\bPhi}{\boldsymbol{\Phi}}
\newcommand{\brho}{\boldsymbol{\rho}}
\newcommand{\btheta}{\boldsymbol{\theta}}
\newcommand{\bvarpi}{\boldsymbol{\varpi}}
\newcommand{\bpi}{\boldsymbol{\pi}}
\newcommand{\bpsi}{\boldsymbol{\psi}}
\newcommand{\bxi}{\boldsymbol{\xi}}
\newcommand{\bzeta}{\boldsymbol{\zeta}}
\newcommand{\bx}{\boldsymbol{x}}
\newcommand{\by}{\boldsymbol{y}}

\newcommand{\cA}{{\cal A}}
\newcommand{\bcA}{\boldsymbol{\cal A}}
\newcommand{\cB}{{\cal B}}
\newcommand{\cC}{{\cal C}}
\newcommand{\cD}{{\cal D}}
\newcommand{\cE}{{\cal E}}
\newcommand{\cG}{{\cal G}}
\newcommand{\cH}{{\cal H}}
\newcommand{\bcH}{\boldsymbol {\cal H}}
\newcommand{\cI}{{\cal I}}
\newcommand{\cK}{{\cal K}}
\newcommand{\cL}{{\cal L}}
\newcommand{\cM}{{\cal M}}
\newcommand{\cO}{{\cal O}}
\newcommand{\cR}{{\cal R}}
\newcommand{\cS}{{\cal S}}
\newcommand{\dcS}{\ddot{{\cal S}}}
\newcommand{\ds}{\ddot{{s}}}
\newcommand{\cT}{{\cal T}}
\newcommand{\cU}{{\cal U}}
\newcommand{\cY}{{\cal Y}}
\newcommand{\wt}[1]{\widetilde{#1}}

\newcommand{\mA}{\mathbb{A}}
\newcommand{\mE}{\mathbb{E}}
\newcommand{\mG}{\mathbb{G}}
\newcommand{\mR}{\mathbb{R}}
\newcommand{\mS}{\mathbb{S}}
\newcommand{\mU}{\mathbb{U}}
\newcommand{\mV}{\mathbb{V}}
\newcommand{\mW}{\mathbb{W}}

\newcommand{\uq}{\underline{q}}
\newcommand{\ubq}{\underline{\boldsymbol q}}

\newcommand{\red}[1]{\textcolor[rgb]{1,0,0}{#1}}
\newcommand{\gre}[1]{\textcolor[rgb]{0,1,0}{#1}}
\newcommand{\blu}[1]{\textcolor[rgb]{0,0,0}{#1}}

\title{Distributed Resource Allocation for Network Slicing over Licensed and Unlicensed Bands}

\author{Yong~Xiao, \IEEEmembership{Senior Member, IEEE}, Mohammed Hirzallah, and Marwan Krunz, \IEEEmembership{Fellow, IEEE}

\thanks{An abridged version of this paper was presented at the IEEE SECON Conference, Hong Kong, China, June 2018\cite{XY2018NetSliceSECON}. This work is partially supported by the National Key R\&D Program of China under Grant No. 2016YFE0133000 and EU Horizon2020 under Grant No. EXICITING-723227.

Y. Xiao is with the School of Electronic Information
and Communications at the Huazhong University of Science and Technology, Wuhan, China (e-mail: yongxiao@hust.edu.cn).  

M. Hirzallah and M. Krunz are with the Department of Electrical and Computer Engineering at the University of Arizona, Tucson, AZ (e-mails: \{hirzallah, krunz\}@email.arizona.edu).} 
}


\maketitle
\begin{abstract}
Network slicing is considered one of the key enabling technologies for 5G due to its ability to customize and ``slice" a common resource to support diverse services and verticals. 
This paper introduces a novel {\em inter-operator network slicing framework} in which multiple mobile network operators (MNOs) can coordinate and jointly slice their accessible spectrum resources in both licensed and unlicensed bands. 
For licensed band slicing, we propose an {\em inter-operator spectrum aggregation method} that allows two or more MNOs to cooperate and share their licensed bands to support a common set of service types.
%
%
We then consider the sharing of unlicensed bands. Because all MNOs enjoy equal rights to accessing these bands, we introduce the concept of {\em right sharing} for MNOs to share and trade their spectrum access rights. 
We develop a {\em modified back-of-the-envelope (mBoE) method} for MNOs to evaluate their {\em Value-of-Rights (VoR)} when coexisting with other wireless technologies. A {\em network slicing game} based on the overlapping coalition formation game is formulated to investigate cooperation between MNOs.
We prove that our proposed game always has at least one stable slicing structure that maximizes the social welfare.
To implement our proposed framework without requiring MNOs to reveal private information to other MNOs, we develop a distributed algorithm called {\em Distributed Alternating Direction Method of Multipliers with Partially Variable Splitting} (D-ADMM-PVS).
Performance evaluation of our proposed framework is provided using a 
discrete-event simulator that is driven by real MNO deployment scenarios based on 
over 400 base station locations deployed by two primary cellular operators in the city of Dublin. 
Numerical results show that our proposed framework can almost double the capacity for 
all supported services for each MNO in an urban setting.  
\end{abstract}



\section{Introduction}
\label{Section_Introduction}
To meet the demand for 5G networks, 
MNOs have taken steps to secure more spectrum resources. 
In particular, the concept of inter-operator spectrum sharing, also referred to as the co-primary spectrum sharing\cite{Singh2015IOSS}, has been used to allow two or more MNOs to share their licensed bands with each other, therefore 
significantly increasing the spectrum instantaneously available to each individual MNO.
Both FCC and 3GPP have recently set forth several initiatives aiming at encouraging spectrum sharing between MNOs. More specifically, 3GPP Release 14 promotes the idea of radio access network (RAN) sharing, which allows multiple MNOs to share network resources, including infrastructure, network functions, and spectrum to improve spectrum utilization and reduce system roll-out cost/delay\cite{3GPP2016NetworkShare}. FCC has introduced new co-primary shared access rules 
for several  millimeter wave (mmWave) bands
to promote cooperation and spectrum sharing among spectrum licensees \cite{FCCmmW2016}.
To further alleviate spectrum scarcity in commercial cellular systems, MNOs have been allowed to extend their services to unlicensed bands, including the 5 GHz unlicensed-national-information-infrastructure (U-NII) radio band\cite{Qualcomm2014UnlicensedLTE} as well as the 57-64 GHz and 64-71 GHz bands, recently opened up by FCC\cite{FCCmmW2016}. Currently, major MNOs such as AT\&T, Deutsche Telekom, and China Mobile, are actively deploying network infrastructure to support cellular services over unlicensed bands\cite{Qualcomm2014UnlicensedLTE}.

In addition to the improvement in transmission speeds and supported traffic volumes, 5G networks are expected to serve highly heterogenous services with diverse quality-of-service (QoS) requirements.
Network slicing is considered a key enabler for 5G, due to its ability to create logical partitions of a common resource such as radio spectrum and/or network infrastructure.
These partitions, known as the network slices, can be orchestrated and customized according to different service requirements. Existing works on network slicing can be classified into two categories: infrastructure slicing and spectrum resource slicing. The former allows a set of common network equipments such as antennas, computing and storage equipments to be ``sliced" into logical networks each of which can be tailored for each specific type of service. The later focuses on partitioning of spectrum resources for supporting different service types. In this paper, we focus on the spectrum resource slicing and, to simplify our description, we use `spectrum resource slicing' and `network slicing' interchangeably.
Network slicing has the potential to significantly improve spectrum efficiency and enable more flexible and novel services with stringent QoS requirements that cannot otherwise be supported by the existing architecture.

One key challenge in inter-operator network slicing that remains relatively unexplored is how to efficiently allocate resources over both licensed and unlicensed bands according to different QoS requirements of different services.
Licensed and unlicensed bands exhibit different characteristics and require different mechanisms to access. In particular, a licensed band is typically allocated to an MNO for exclusive use. MNOs have already carefully planned their network infrastructure and adopted various centrally controlled resource scheduling and allocation mechanisms to ensure optimal utilization and reliable service support for user equipments (UEs). The unlicensed band, on the other hand, is open to many different wireless technologies. To reduce contention between coexisting systems, Wi-Fi and licensed assisted access (LAA) LTE standards rely on a carrier sense multiple access (CSMA) mechanism, commonly referred to as listen-before-talk (LBT). In this mechanism, both LAA and Wi-Fi transmitters must first sense the channel and can only access it if it is deemed to be idle. The uncertainty in the channel access delay in unlicensed bands makes it difficult to support services with stringent QoS requirements. Therefore, most existing works\cite{Leconte2018NetSlice, Sciancalepore2017NetSlice, Caballero2017NetSlicingGame} focus on network slicing focusing on licensed bands. How to share and jointly slice the unlicensed spectrum among MNOs is still an open problem.

In this paper, we address the above challenge by designing a novel framework that allows multiple MNOs to jointly distribute and orchestrate licensed and unlicensed spectrum resources according to the service demands and requirements of their UEs. More specifically, for licensed band slicing, we propose an {\em inter-operator spectrum aggregation} method that allows two or more MNOs to access each other's licensed spectrum. In this method, an MNO divides its licensed band into partitions, each of which is intended to support a specific type of service (i.e., a given set of QoS values). Multiple MNOs can then aggregate their distributed licensed bands to support traffic associated with the same type of service.
We introduce the concept of {\em right sharing} to investigate inter-operator cooperation over unlicensed bands. According to this concept, each MNO will first quantify the benefit that can be obtained from operating on the unlicensed band, referred to as the {\em Value-of-Rights (VoR)}. MNOs can then negotiate and trade their rights to access unlicensed bands according to the estimated value. We propose a {\em modified back-of-the-envelope (mBoE)} method for each MNO to estimate its VoR as well as the potential performance improvement that can be gained when one or more other MNOs are willing to give up their rights for accessing the unlicensed band.
We observe that if each MNO is given the choice to slice both licensed or unlicensed bands, the interaction between MNOs can be very complex. For example, if an MNO cannot secure enough 
licensed spectrum, it becomes more aggressive and willing to pay more to other MNOs to reduce contention over the unlicensed spectrum. Similarly, if the licensed spectrum can offer sufficient resources to support the required traffic of an MNO, this MNO will have more incentive to sell its right over the unlicensed bands to other MNOs. To investigate the interaction among MNOs, we develop a  {\em network slicing game} based on the overlapping coalition formation game.
In this game, MNOs can jointly decide the spectrum allocation as well as distribution of the utility obtained in each network slice.
A network slicing structure can only result in a stable state when no MNO can benefit from unilaterally deviating from this structure. It is known that analyzing an overlapping coalition formation game is notoriously difficult. Such a game does not always admit a stable structure. Furthermore, allowing overlaps between coalitions, e.g., each player instead of being a member of a single coalition can be associated with multiple coalitions, results in infinitely many possible structures, which makes exhaustive search-based methods, widely used in typical partition-based coalition formation games, impossible to apply. We prove that our proposed network slicing game always admits at least one stable structure. 

We observe that the existing centralized network slicing architecture cannot be directly extended to the inter-operator scenario.
Accordingly, we develop a novel {\em Distributed Alternating Direction Method of Multipliers with Partial Variable Splitting} (D-ADMM-PVS) algorithm to implement our proposed network slicing in a distributed manner. D-ADMM-PVS does not require back-and-forth exchange of private information among MNOs.
We prove that our proposed algorithm can approach the stable and optimal network slicing structure in linear time.
Performance evaluation of our proposed framework is provided using a 
discrete-event simulator that is driven by real MNO deployment scenarios based on 
over 400 base station locations deployed by two primary cellular operators in the city of Dublin. 
Numerical results show that our proposed framework can almost double the capacity for 
all supported services for each MNO under the urban environment.

To evaluate the practical performance of our proposed framework, we consider the actual base station (BS) topological deployment made by two major telecommunication operators in Ireland coexisting with the Wi-Fi APs installed at all the Starbucks coffee shops in the city of Dublin. We consider the scenario that all the cellular BSs have been upgraded to support LAA LTE operations. We develop a C++-based discrete-event simulator using CSIM development toolkit\cite{CSIM_lib} to simulate the possible contention between LAA BSs and Wi-Fi APs.
Our numerical results show that our proposed framework can almost double the capacity for 
all supported services in the urban scenario even when only two MNOs cooperate.

\section{Related Work}
\label{Section_Background}

\noindent
\blu{\bf Inter-operator Spectrum Sharing in Licensed Band:} Most existing work on inter-operator spectrum sharing focuses on licensed band sharing between MNOs with similar traffic characteristics  and spectrum allocations. More specifically, European Commission's Mobile and wireless communications Enablers for Twenty-twenty Information Society (METIS) future spectrum system concept \cite{Koufos2015METISIOSS} suggests two scenarios for licensed band sharing between MNOs: 
limited spectrum pooling (LSP) and mutual renting (MR). In LSP, two or more MNOs contribute part of their licensed spectrum to form a common pool\cite{XY2013SpectrumPool}. All contributing MNOs have equal rights to access the pool and should follow a mutually agreed upon rule to access the pooled resource.
MR allows each MNO to 
temporally license part or all of its spectrum to another MNO. In contrast to LSP, 
each MNO in MR can maintain a strict access priority over its own licensed band\cite{XY2015IOCA}. Inter-operator network sharing was also recently adopted by 3GPP, where multiple MNOs can jointly manage the commonly shared resource\cite{3GPP2016NetworkShare}.
In \cite{Guo2014SpectrumCooperation}, the authors studied the energy-efficient spectrum cooperation between different cellular systems with the objective of reducing MNOs' operational costs. The authors in \cite{Duan2011Operator} investigated the optimal investment and pricing decision of a cognitive mobile virtual network operator under uncertain spectrum supply.

\begin{figure}
\centering
\includegraphics[width=2.5 in]{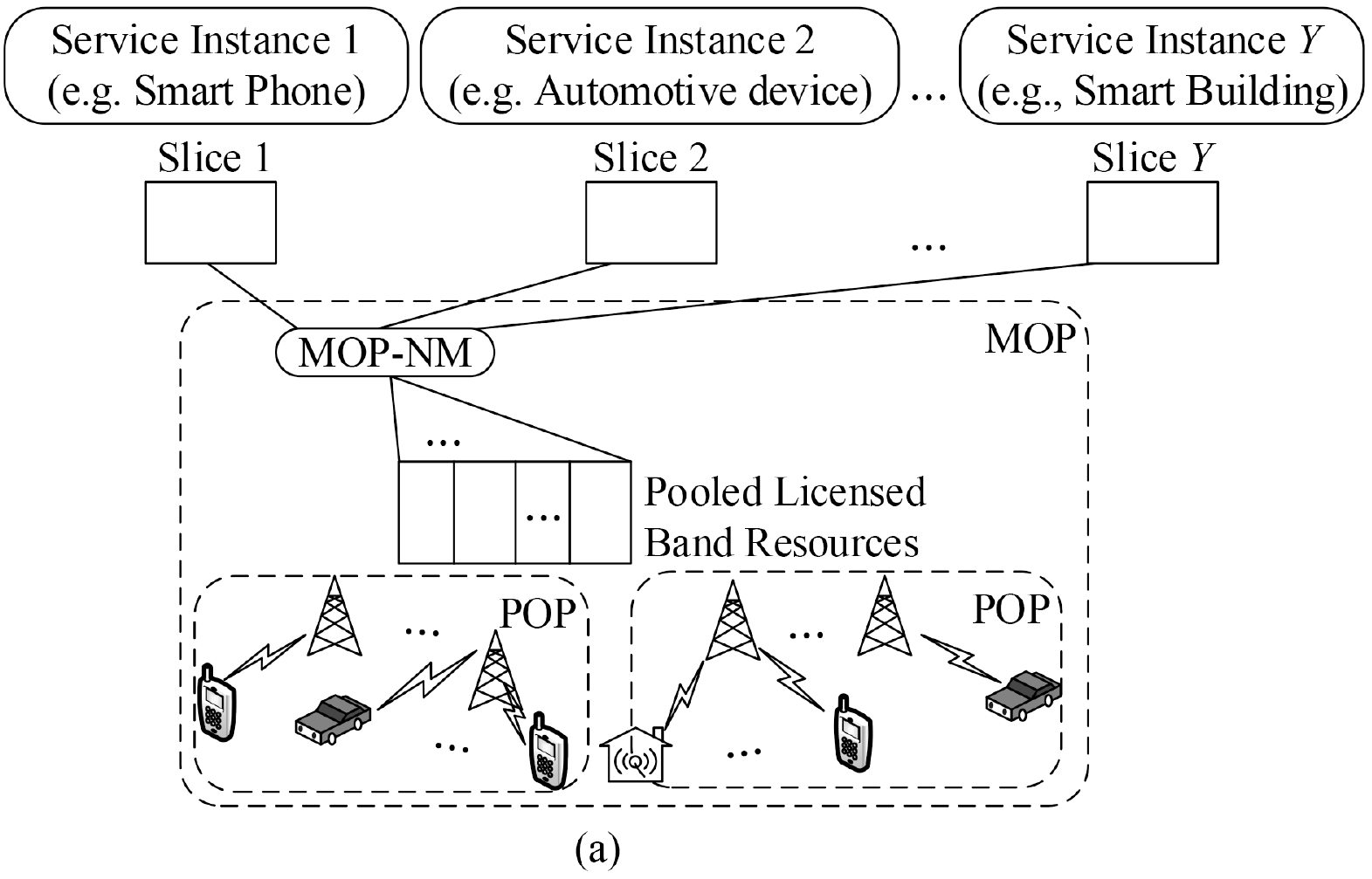}
\includegraphics[width=2.5 in]{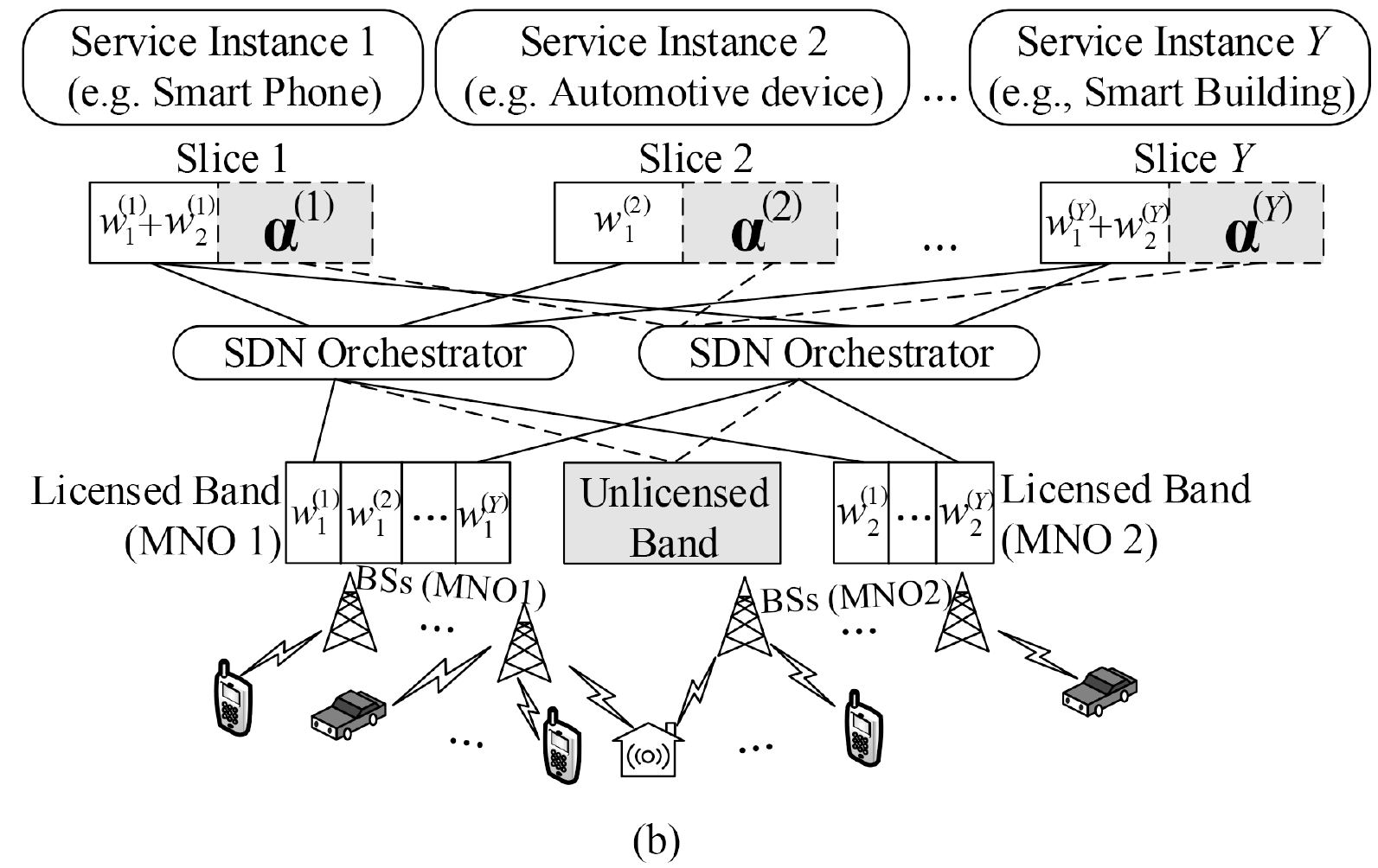}
\vspace{-0.1 in}
\caption{(a) 3GPP's active network sharing management architecture and (b) our proposed inter-operator network slicing framework.}
\vspace{-0.4in}
\label{Figure_BloackDiagram}
\end{figure}

\noindent
\blu{\bf Licensed Band Network Slicing: } Inter-operator resource sharing has recently been studied from the network slicing perspective\cite{Caballero2017NetSlicingGame, Sciancalepore2017NetSlice, Leconte2018NetSlice}. In particular, in \cite{Caballero2017NetSlicingGame}, a resource allocation mechanism called the Fisher market was used to study resource allocation across slices. In \cite{Sciancalepore2017NetSlice}, a signaling-based network slicing broker solution was proposed to achieve accurate traffic prediction, slice scheduling, and admission control. The authors in \cite{Leconte2018NetSlice} proposed a fine-grained resource allocation for slices of licensed spectrum both in terms of bandwidth and cloud processing. \blu{In \cite{Bagaa2018vCoreNetSlice}, the authors investigated the problem of VNF placement. Two algorithms have been proposed. The first one calculates the optimal number of virtual instances of 4G (MME, SGW, and PGW) or 5G (AMF, SMF, and AUSF) core network elements and the second one derives the placement of these virtual instances over a federated cloud. In \cite{Afolabi2017E2ENetSlic}, the authors carried out series of experiments involving the deployment of different VM flavors to determine the proper 
amount of resources that are most suitable to host a virtualized evolve packet cores (vEPCs) considering the business requirements it has to accomplish. In \cite{Taleb2017NetworkSlic5G}, the authors assessed the potential of network slicing to provide
the appropriate customization and highlights its technology challenges. In \cite{Afolabi2017NetSlic}, the authors introduced a 5G architecture that enables the orchestration, instantiation and management of end-to-end network slices over multiple administrative and technological domains. For more detailed survey of network slicing in 5G network architecture, we refer the reader to \cite{Taleb2017NetSlic, Afolabi2018NetSlic} and the references therein. }

\noindent
\blu{\bf Inter-operator Spectrum Sharing in Unlicensed Band: }
Compared to the resource sharing over licensed bands, the sharing of unlicensed spectrum is much more complicated due to the heterogeneity of coexisting systems and technologies. In \cite{Teng2017UnlicensedSS}, the authors studied a scenario in which the unlicensed band can be divided into several partitions, each of which can be exclusively accessed by one MNO. A spectrum sharing scheme was proposed to allow spectrum borrowing and lending among MNOs.
Motivated by recent observations that Wi-Fi and LTE coexistence in the unlicensed band could result in more than 70\% throughput degradation for Wi-Fi systems\cite{Guan2016LTEWiFi, Abdelfattah2017LTEWiFi}, many existing works focused on developing mechanisms to ensure fair coexistence between LTE and Wi-Fi\cite{Hasan2016LAA}. For example, the authors in \cite{Hasan2016LAA} proposed to adjust the contention parameters  of LAA to achieve a fair coexistence between Wi-Fi systems and co-locating LAA cells.

\section{Distributed Network Slicing Framework}
\label{Section_Framework}
In this section, we introduce a distributed network slicing framework
that allow multiple MNOs to decide the slicing of their accessible spectrum resources in both licensed and unlicensed bands. We consider a set of $Y$ types of services, labeled as ${\cal Y} = \{1, 2, \ldots, Y\}$, to be supported by all MNOs.
%
To manage and orchestrate the spectrum slicing over multiple MNOs with a diverse set of services,
we introduce a software-define networking (SDN)-based framework, in which the control functions
are implemented by a set of programmable network entities referred to as the SDN orchestrators each of which can coordinate the resource allocation among the network infrastructures within a specific coverage area. Our framework extends from 3GPP's active network sharing management architecture\cite{3GPP2016NetworkShare}. 
In the 3GPP's architecture, a master operator (MOP) deploys a so-called MOP's network manager (MOP-NM), a centralized SDN-controller to collect global information and manage the allocation of the shared radio resource among multiple participating operators (POPs), as illustrated in Figure \ref{Figure_BloackDiagram}(a).
This centralized management approach cannot be directly applied to optimize a large-scale inter-operator slicing system due to the following reasons. First, in the 3GPP's architecture, MOP-NM monitors and controls a fixed amount of resource (e.g., network infrastructure and/or spectrum) shared among a fixed set of POPs. However, in practice, different MNOs can have different demands and requirements for different services. In this scenario, each MNO may like to cooperate with different subsets of MNOs to support different types of services.
Second, the 3GPP's architecture is limited to sharing licensed band resources.
Compared to licensed bands, unlicensed bands are open to all and contain much wider bandwidth for MNOs to access.
At the same time, unlicensed bands require different spectrum access mechanisms than those used for licensed bands.
Third,  POPs can share network infrastructure and resources across a wide geographical area. Allowing the MOP-NM to always collect global information from all POPs will lead to network congestion and intolerably high latencies. Fourth,
some POPs may not want to disclose their private proprietary information to the MOP.

To address the above issues, we introduce a distributed framework based on a set of SDN orchestrators that can be distributed throughout a large service area.
An orchestrator can reside in a serving gateway (S-GW) of a single MNO or a gateway core network (GWCN) share among multiple MNOs.
If an MNO decides to jointly slice resources with other MNOs to serve UEs in a specific coverage area, it will coordinate through the local SDN orchestrator to decide how much spectrum resource to be distributed for each type of service before creating service instances. The orchestrator can only access the core network elements such as Mobility Management Entity (MME) and Package Data Gateways (P-GW) e.g., via S1 interface, of a group of MNOs when a contract has been mutually agreed upon.
In our framework, MNOs are self-interested and will only cooperate with each other if all the MNOs can benefit from cooperation and also obtain a fair share of the total revenue. We focus on distributed optimization of resource slicing among MNOs according to different service requirements taking into account the different spectrum and channel access mechanisms in licensed and unlicensed bands.

\section{Inter-operator Network Slicing}
\label{Section_Slicing}

In this section, we first propose the inter-operator network slicing for licensed bands in Section \ref{Subsection_SlicingLicensedBand}. For network slicing over unlicensed bands, we introduce the concept of VoR for each MNO in Section \ref{Subsection_SlicingUnlicensedBand}. Finally, we formulate the joint network slicing problem over licensed and unlicensed bands in Section \ref{Subsection_SlicingLicensedUnlicensedBand}.

\subsection{Network Slicing for Licensed Bands}
\label{Subsection_SlicingLicensedBand}
We consider a wireless system consisting of a set of $M$ MNOs, labeled as ${\cal M} = \{1, 2, \ldots, M\}$. Each MNO $i$, $i\in {\cal M}$, can provide services through its network infrastructures, e.g., BSs, 
$B_i$ Hz bandwidth of licensed spectrum.
\blu{Each MNO can offer at most $Y$ types of service each of which 
is associated with a specific QoS requirement, e.g., a minimum throughput that must be guaranteed. In this paper, we consider systems under backlogged traffic such that each UE always generates saturated traffic for all supported service types. This assumption is reasonable in the sense that previous study shows that only limited performance improvement can be achieved by network slicing when a limited service traffic has been requested. 
Let $\eta^{(l)}_i$ be the minimum throughput that is required by type $l$ service offered by MNO $i$.}

Each MNO can divide and aggregate the contiguous and uncontiguous parts of the licensed band to support various types of services.
Each MNO $i$ divides $B_i$ frequency band into a set of subcarriers, each of which can be allocated to support a particular service type. Generally speaking, the bandwidth of a subcarrier is much smaller than $B_i$. For example, in LTE, the minimum bandwidth for each subcarrier is 15 kHz and the maximum bandwidth that can be allocated to an individual service is 100MHz with carrier aggregation ~\cite{LTEA-36.912}. The bandwidth that can be accessed by each MNO will be higher if two or more MNOs can aggregate their licensed band together using 3GPP network sharing architecture\cite{3GPP2016NetworkShare}.  
We can, therefore, assume the licensed band is continuously dividable among different types of service.

Instead of accessing its own licensed band, each MNO can also negotiate with other MNOs to form a group for possible sharing of the licensed bands. We refer a group of MNOs that decide to share their licensed bands with each other for supporting type $l$ service as a Service Support Group (SSG) denoted as ${\cal C}^{(l)}$ for $\cC^{(l)} \subseteq {\cal M}$.

Let $w^{(l)}_{i}$ be the portion of licensed band distributed by MNO $i$ to support the $l$th type of service. We have $0 \le w^{(l)}_{i}\le B_i$. MNOs in ${\cal C}^{(l)}$ will aggregate their allocated licensed bands for type $l$ service traffic. We can then write the total aggregated licensed spectrum allocated by MNOs to support type $l$ service as $w^{(l)} = \sum_{i \in {\cal C}^{(l)}} w^{(l)}_i$. Let ${\cal L}_i$ be the set of all the communication links associated with UEs of MNO $i$. We write $d^{(l)}_{k,i}$ as the portion of $w^{(l)}$ that can be accessed by the $k$th communication link (e.g., uplink or downlink from each UE or BS) to send data traffic associated with type $l$ service, \blu{i.e., the total spectrum that can be accessed by each link $k$ of MNO $i$ is given by $u^{(l)}_{k,i} = d^{(l)}_{k,i} w^{(l)}$ for $\sum_{k\in {\cal L}_i, i\in {\cal M}} u^{(l)}_{k,i} \le w^{(l)}$. Each link associated with a member MNO will follow a mutually agreed scheduling procedure to access the aggregated spectrum. The final portion of aggregated spectrum that can be accessed by each link will also depend on the specific network topology as well as traffic from other nearby UEs. For example, suppose $n$ links co-located in the same area send requests for the same type $l$ service at the same time. A commonly adopted approach is to equally allocate the aggregated spectrum to each service requesting link, i.e., we have $d^{(l)}_{k,i} = {1 \over n}$ and each link $k$ is allocated with $u^{(l)}_{k,i} = {w^{(l)} \over n}$ bandwidth of the aggregated spectrum. In this paper, we consider a generalized framework in which each MNO can optimize vector $\bu_i = \langle u^{(l)}_{k,i} \rangle_{k\in {\cal L}_{i}, l \in {\cal Y}}$ consisting of the allocated spectrum for each type of service at each link. More specifically, we can write the utility obtained by MNO $i$ for serving type $l$ service at the $k$th link as $\pi^{(l)}_{k,i} \left( u^{(l)}_{k,i} \right) = \rho^{(l)}_{i} u^{(l)}_{k,i} R_{k,i}$
where $\rho^{(l)}_i$ is the price per data bit charged by MNO $i$ by serving type $l$ service and $R_{k,i} = \log_2 \left( 1 + {\rm SNR}_{k,i} \right)$ is the throughput per unit (Hz) achieved by link $k$ and ${\rm SNR}_{k,i}$ is the received signal-to-noise ratio for link $k$ when it is the only link to access the channel.}

If two or more MNOs perform network slicing by jointly sharing their licensed bands, we can write the optimization problem for each MNO $i$ as follows:
\begin{subequations}\label{eq_LicensedSlicing}
\begin{align}
& \max\limits_{\bu_i = \langle u^{(l)}_{k,i} \rangle_{k\in {\cal L}_{i}, l \in {\cal Y}}}  \sum_{k \in {\cal L}_i} \sum\limits_{l\in {\cal Y}} \pi^{(l)}_{k,i} \left( u^{(l)}_{k,i} \right) \label{eq_LicensedSlicing_Obj} \\
&\;\;\;\;\;\;\;\;\; {\rm s.t.} \;\;\;\;\; \sum\limits_{l\in{\cal Y}} u^{(l)}_{k,i} \le \sum\limits_{j\in {\cal C}^{(l)}} B_{j} \mbox{ and } R_{k,i} 
u^{(l)}_{k,i} \ge \eta^{(l)}_i, \nonumber \\
& \;\; \forall k \in {\cal L}_i, l\in {\cal Y}, \label{eq_LicensedSlicing_Con1}
\end{align}
\end{subequations}
where the first term in (\ref{eq_LicensedSlicing_Con1}) means that the combined licensed spectrum allocated to each link for supporting all types of service cannot exceed the total available bandwidth owned by the resource sharing MNOs. And the second term in in (\ref{eq_LicensedSlicing_Con1}) specifies the minimum throughput that must be guaranteed for each supported type of service.

The licensed band network slicing can be directly implemented by the co-primary spectrum shared access specified in METIS' future spectrum system concept with LSP mode\cite{Singh2015IOSS}. In particular, if MNOs in an SSG decide to operate in the LSP, all the member MNOs will negotiate for a group license and use the inter-operator carrier aggregation strategy proposed in \cite{XY2015IOCA}  to form a common resource pool $w^{(l)}$ 
to support type $l$ service.

\subsection{Network Slicing for Unlicensed Band}
\label{Subsection_SlicingUnlicensedBand}

Compared to the sharing of licensed band in which each MNO has an exclusively licensed resource that can be traded with others, in the unlicensed band, all MNOs enjoy the same right to access the same amount of spectrum resource. How to develop a framework that can incentivize the sharing among MNOs for their spectrum access rights in the unlicensed band is still an open problem.

We propose the concept of {\em (spectrum access) right sharing} for inter-operator network slicing in unlicensed band. In this concept, each MNO can share its spectrum access right with other MNOs. Before we formally introduce the concept, we need to first characterize the {\em value-of-rights (VoR)} for each MNO in the unlicensed band. We have the following observations:
\begin{itemize}
\item[C1)] The VoR of an MNO in unlicensed band depends on its benefit that can be obtained in the unlicensed band.
Different MNOs can observe different benefits in unlicensed band. The MNOs that can obtain higher benefits in the unlicensed band will less likely to give up their rights compared to others.

\item[C2)] The compensation that can be provided by an MNO $i$ for another MNO $j$ to give up the right to access unlicensed band is closely related to the benefit improvement that can be obtained by MNO $i$ when the UEs and the BSs associated with MNO $j$ stop accessing the unlicensed band. In other words, some MNOs are willing to pay higher prices than others for the right of a certain MNO.   

\item[C3)] When an individual MNO stop accessing the unlicensed band, all the other co-locating MNOs can benefit from the reduction of channel contending UEs and BSs.
\end{itemize}

From C1) and C2), we can observe that it is important for each MNO to first pre-evaluate its benefit that can be obtained in the unlicensed band. From C3), we can observe that it makes sense to let all the beneficial MNOs to cooperate and jointly compensate 
to  the MNOs that are willing to stop accessing the unlicensed band.

Let us now introduce the inter-operator right sharing framework in unlicensed band.

\subsubsection{LAA Protocol}
Before we discuss the right sharing in unlicensed bands, let us first briefly review the CSMA-based LAA protocol. Since the unlicensed band is open to all wireless technologies, to avoid the collision and cross-interference, data transmission is required to follow an LBT-based channel access mechanism. In this mechanism, each UE or BS must first sense the vacancy of the channel for a duration of time called distributed inter-frame spacing (DIFS). If the channel is sensed busy, the UE or BS will defer the transmission until the channel becomes idle and then wait for a DIFS duration plus a random number, referred to as the backoff counter number, of time slots before accessing the channel. The value of the backoff counter is uniformly randomly generated between 0 and an integer value called contention window $CW$. The backoff counter is decremented one-by-one for each time slot till zero when the channel is idle. In case that the channel is occupied by other neighboring UE or BS, the backoff counter will be frozen until the channel is sensed to be idle again.

As observed from the above description, it is generally impossible to guarantee the successful channel access in the unlicensed band, e.g., even the probability of channel access is high, there is still a small chance that an LTE UE or BS cannot send any data packet on the unlicensed band.

Let $\xi_{k,i}$ be the probability of channel access for the $k$th link associated MNO $i$. We also use $B^{(u)}$ to denote the total amount of spectrum resource of the unlicensed band.

\subsubsection{Estimation of Probability of Access in Unlicensed Band}
Before negotiating with other MNOs, an MNO needs to first pre-evaluate the potential benefit that can be obtained in the unlicensed band. It also needs to identify whether to negotiate with one or more other MNOs for the possibility of giving up their rights in the unlicensed band. 
Similarly, once an MNO receives a request from other MNOs about the possibility to give up its right to access the unlicensed band, it needs to estimate how much damage it will cause and how much compensation it should expect from the requesting MNOs. In this paper, we assume the benefit of each MNO in unlicensed bands is closely related to its probability of channel access for its BSs and UEs. 

We introduce an mBoE method for each MNO to pre-evaluate the probability of access for each of its links. The basic idea is to generate a graphical model that can characterize the possible contention among all the intra- and inter-operator links as well as the channel contentions from other coexisting wireless technologies such as Wi-Fi. Our mBoE method is extended from the original back-of-the-envelop (BoE) method previously introduced in \cite{Liew2010BoE}. BoE is a simple and effective method that can quickly calculate the probability of access of a contention graph without requiring any detailed information about locations and transmission parameters. 

Unfortunately, the original BoE cannot be directly applied into LAA system due to the following reasons: 1) the original BoE method was built on a homogeneous 802.11 network in which all the devices have the same contention parameters. In our system, the LAA BSs and UEs coexist with other wireless technologies such as Wi-Fi, 2) the BoE method needs to have a complete contention graph consisting of all the communication links and the calculation of each link requires to consult the entire network topology. 
However, in our muti-MNO system, each MNO cannot know the relative locations of UEs or BSs associated with other MNOs. To address these two issues, our mBoE is built on an empirical table consisting of the pre-measured probability of access of each LAA BS and UE when contending with different subsets of Wi-Fi and/or other LAA transmitters under different network topologies. Compared to the original BoE method, our proposed mBoE method provides an improved estimation results with reduced computation complexity. In addition, our mBoE can calculate the probability of access for each local link using only the local network topology.

Before we introduce the detailed method, let us introduce the following assumptions. Note that these assumptions are only introduced for justifying the mBoE method and are not necessary for our network slicing game or distribution algorithms introduced later in the paper.
\begin{itemize}
\item[A1)] Each UE or BS can sense the transmission of neighboring UEs and BSs which can be associated other MNOs as well as Wi-Fi transmitters,
\item[A2)] The time duration for random backoff countdown can be considered as negligible, compared to the duration spent on data transmission,
\item[A3)] The probability distributions of the long-term residual backoff countdown time and data transmission time are stationary.
\end{itemize}

\begin{table}[tbp]
\centering
\caption{Wi-Fi and LAA channel access parameters\cite{3GPP2016LAA}}
\vspace{-0.1in}
\label{Tabel_LAAandWiFi}
\begin{tabular}{lllll}
\hline
 & DIFS & $CW_{\min}$ & $CW_{\max}$ & TXOP  \\
\hline
802.11ac & 34 ms & 3 & 7  & 1.504ms \\
\hline
LAA & 25 ms & $3$ & $7$ & $2$ msec \\
\hline
\end{tabular}
\vspace{-0.2in}
\end{table}
\normalsize

\blu{Assumption A1) is adopted by many existing works and is commonly considered as reasonable because both LAA and Wi-Fi use OFDM to send data signal which is known to contain a periodically repetitive signal to be sent for mitigating inter-symbol interference (ISI), channel estimation, and network synchronization purpose. More specifically, 
LTE and Wi-Fi transmit signals, like any other OFDM modulated signal, adopt the concept of cyclic prefix (CP) to mitigate ISI between two consecutive symbols. The CP is a replication of the end of an OFDM symbol that is copied and added to the beginning of that symbol. Each LAA UE and BS can estimate the correlation between consecutive data signals and detect the time duration (TXOP duration) for each channel occupancy with a simple sliding-window approach\cite{Samy2018WiSec}. 
In addition to the CP-based approach, the UEs and BSs can also sense the preamble, a known data sequence sent at the beginning of each Wi-Fi frame for channel estimation as well as the LTE reference signals periodically sent by each LTE transmitter to support synchronization to detect their neighboring transmitters. See Table \ref{Tabel_LAAandWiFi} for a list of transmission parameters of LAA release 13\cite{3GPP2016LAA} and 802.11ac Wi-Fi standard\cite{WiFi80211ac}.
}

Note that due to the channel fading and shadowing effects, there may exist the so called ``hidden nodes'', i.e., some BSs or UEs cannot always successfully detect the transmission of their neighboring devices. Since our mBoE method is built on an empirical probability of access table obtained from previous measurements, the effect of the hidden nodes has already been captured in the measuring results. The impact of the hidden nodes can be further reduced by allowing MNOs who have the intention to cooperate to share their sensing results with each other. Each MNO can also extract the local information about channel contenting Wi-Fi transmitters from the beacon signal broadcasted by the connected Wi-Fi APs to further improve its sensing accuracy. 
Assumption A2) follows the same observation as \cite{Liew2010BoE}. In particular, the countdown time of different links may occur concurrently which in some sense cancels the total amount of time spent on resolving the possible collisions among channel contending links. 
We implement the most recent LAA specification in \cite{3GPP2016LAA} into our CSIM-based simulator and our experimental results also verify this observation. In other words, the random backoff mechanism introduced in the CSMA protocols can successfully avoid the collision among channel contending devices for most of the time and therefore in our measuring results, the data transmission time dominates the channel access time.
%
%
Assumption 3) has been proved in \cite{Bianchi2005WiFi} and also verified in our recent work\cite{Samy2018WiSec}. 

The first step of mBoE method is to establish a contention graph that can capture all the contention between the coexisting devices for each MNO.  We formally define contention graph as follows.
\begin{definition}
A {\em contention graph} for a multi-operator LAA system coexisting with other technologies such as Wi-Fi in the unlicensed band is a graph ${\cal G} = \langle {\cal V}, {\cal E} \rangle$ comprising a set ${\cal V}$ of vertices corresponding to the set of all the coexisting links connecting UEs and BSs associated with all the MNOs as well as the coexisting Wi-Fi links and a set $\cal E$ of edges each of which connects two vertices that can sense the existence of each other. 
We also define the contention subgraph associated with MNO $i$ as the subgraph ${\cal G}_i$ of ${\cal G}$ comprising subsets of vertices and edges corresponding to communication links that are only associated with MNO $i$ as well as their sensed entities of other MNOs and Wi-Fi systems.
\end{definition}

Note that it has been verified in \cite{Liew2010BoE} that if two or more contending links are associated with the same technology (e.g., LAA) coexisting with the same set of neighboring devices, the probability for each link to successfully access the channel can be considered as equal. 
If two contending links belong to different technologies, e.g., one link is a LAA link and the other is a Wi-Fi link, then the probability of access must be pre-calculated and pre-stored in a table.

\begin{figure}
\centering
\includegraphics[width=3.5 in]{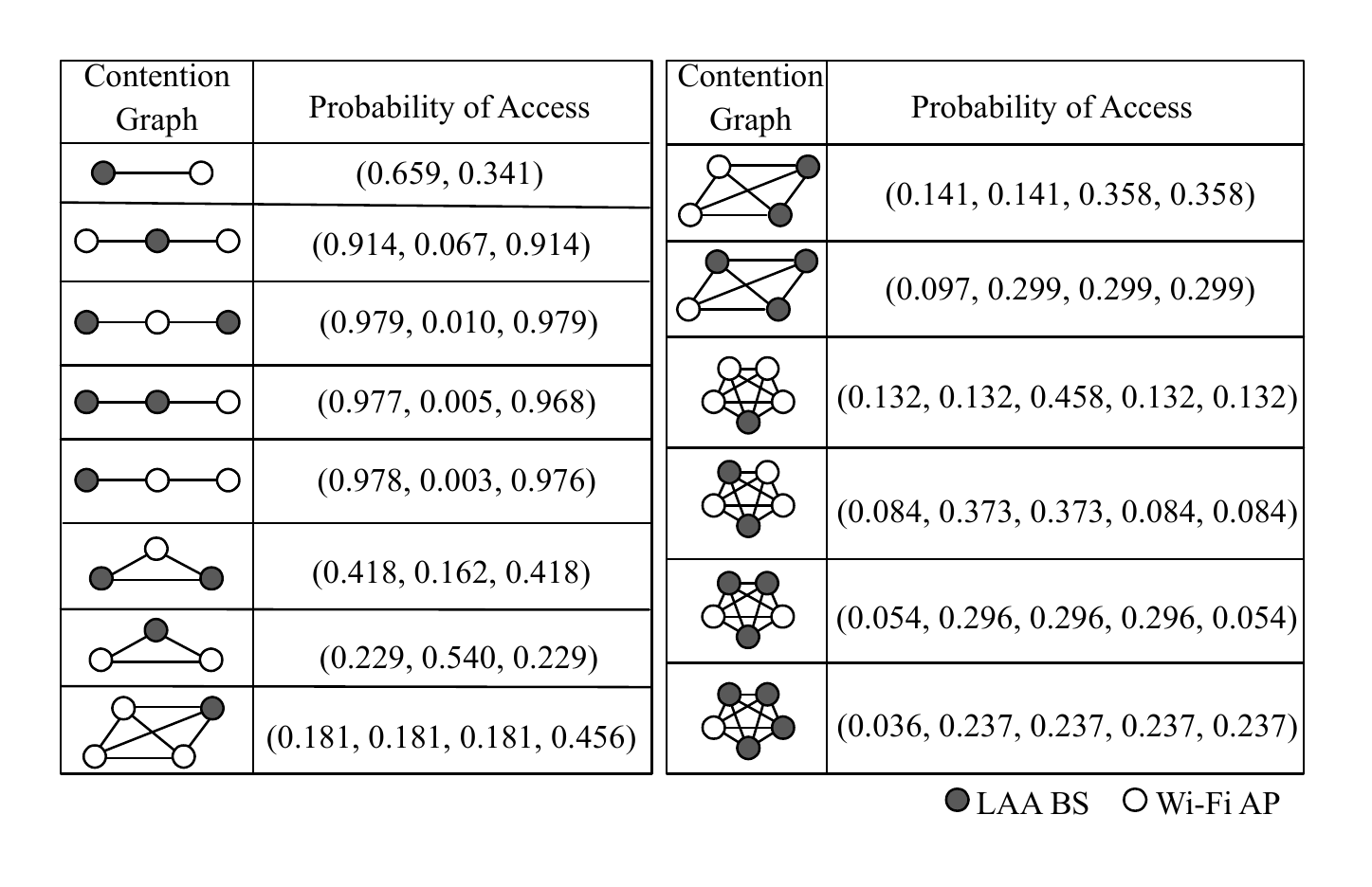}
\vspace{-0.3in}
\caption{Table of contention subgraphs and the corresponding probability of access measured by our CSIM simulator.}
\label{Figure_mBoE}
\end{figure}

In Figure \ref{Figure_mBoE}, we have listed the measured average probability of access under different contention topologies using our developed CSIM-based simulator. Note that since the LAA Release 13 only supports downlink transmission in unlicensed bands, the number of possible contention graphical topologies that involves each BS should be limited, e.g., if all the LAA transmitters correspond to BSs deployed by MNOs, the maximum number of BSs contending with each other in each local area will be equivalent to the number of MNOs. In addition, as observed in many existing results that as the number of channel contending devices becomes large, the probability of channel access will drop significantly. It is therefore unnecessary for each BS to maintain a table that includes a large number of coexisting devices.

Note that in Figure \ref{Figure_mBoE} we observe that the probability of access for a Wi-Fi AP is always lower than that of the LAA BS. This is because we have adopted the most recent LAA specification in Table \ref{Tabel_LAAandWiFi} in which the LAA BS has a shorter DIFS waiting time as well as longer TXOP transmission duration compared to these parameters in 802.11ac Wi-Fi standard. This observation is consistent with the result reported in \cite{Bianchi2005WiFi}.

We define the {\em maximum independent set} for MNO $i$ as follows:
\begin{definition}
An {\em independent set} associated with MNO $i$ is a set of vertices in ${\cal G}_i$ in which no two of which are adjacent. A {\em maximum independent set} for MNO $i$ is an independent set with the largest possible size for graph ${\cal G}_i$.
\end{definition}

It is known that the maximum independent sets of a given graph can be found by standard approaches in polynomial time\cite{Luby1986MaxIndSets}. 
%

One of the main idea behind our proposed procedure is that the maximum independent sets dominate the possible channel contention as well as channel access among all the entities associated with different MNOs coexisting in the same area. In particular, the following proposition has been proved in \cite{Liew2010BoE}.
\begin{proposition}\cite[Propositions 1]{Liew2010BoE}
A CSMA-based system spends most of its time in the maximum independent sets and very little time in other states. 
\end{proposition}

We write the vector for the probability of access for all links associated with MNO $i$ as $\bxi_i = \langle \xi_{k,i} \rangle_{k \in {\cal L}_i}$. 

Each MNO can then use the following procedure to estimate the probability of access for each of its links:
\begin{itemize}
\item[P1)] Establish a contention subgraph ${\cal G}_i$ in unlicensed band using the sensing results from the UEs and BSs of MNO $i$,
\item[P2)] Each MNO $i$ can then identify the possible maximum independent sets of ${\cal G}_i$ using standard approaches,
\item[P3)] Each MNO $i$ generates a modified subgraph ${\cal G}'_i$ by removing all the vertices that are not associated with any maximum independent set from ${\cal G}_i$,
\item[P4)] Each MNO $i$ searches for the probability of channel access $\xi_{k,i}$ for each link $k$ from the pre-stored contention subgraph table.
\end{itemize}

Since each MNO can detect the contention from other MNOs, it can also estimate the possible improvement of the channel access probability if one or more other MNOs stop accessing the unlicensed band. We define the estimated contention subgraph ${\cal G}_{i \backslash j}$ for MNO $i$ as a subgraph of ${\cal G}_i$ such that 
all vertices associated with links from MNO $j$ are removed for $i\neq j$. By replacing graph ${\cal G}_i$ with subgraph ${\cal G}_{i \backslash j}$ in procedure P1), MNO $i$ can re-estimate the resulting probability of channel access $\xi_{k,i \backslash j}$ for each of its links following procedures P2) to P4). Let $\bxi_{i \backslash j} = \langle \xi_{k,i \backslash j} \rangle_{k\in {\cal L}_i}$be the vector of channel access probabilities for all the links associated with MNO $i$ when BSs of MNO $j$ stops accessing the unlicensed band  for $i\neq j$.

\begin{figure}
\centering
\includegraphics[width=3.5 in]{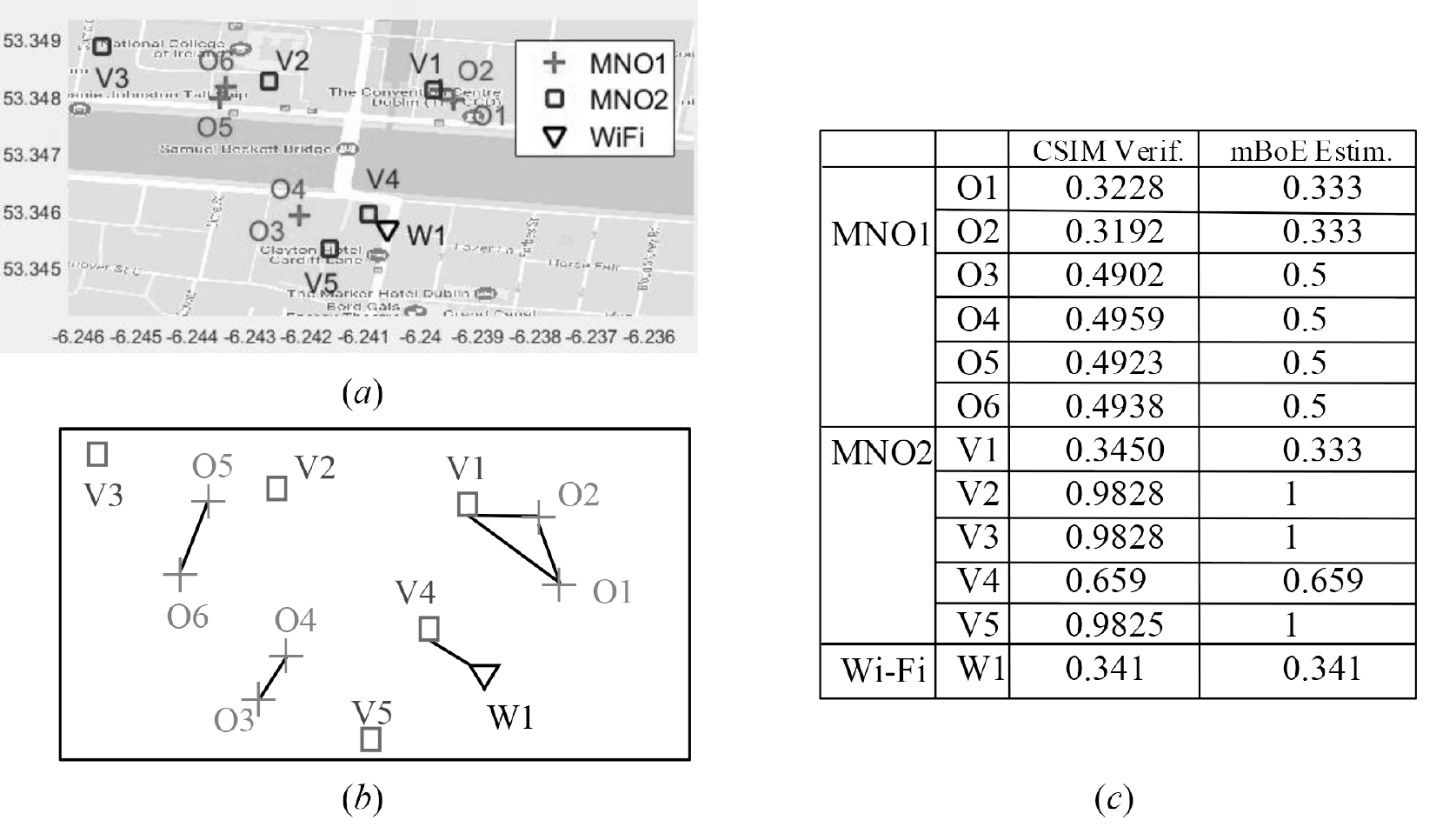}
\vspace{-0.2in}
\caption{Use mBoE to calculate the probability of access in an urban environment: (a) real locations of the BSs, (b) abstracted contention graph, and (c) channel access probability estimated from our proposed mBoE method compared with the real probabilities generated by our CSIM simulator.}
\label{Figure_mBoEExample}
\vspace{-0.2in}
\end{figure}

To verify the performance of our proposed mBoE method, we consider an urban region in the city of Dublin consisting of BSs deployed by two major MNOs in Ireland as well as Wi-Fi APs installed at Starbucks coffee shops in Figure \ref{Figure_mBoEExample}. In particular, we calculate the probability of access estimated from the procedures P1)--P4) and compare these results with the real channel access probability obtained from our developed CSIM simulator. Our result shows that the proposed mBoE can successfully estimate the probability of channel access for each MNO.

\subsubsection{Inter-operator Right Sharing}

As mentioned previously, the performance degradation of an MNO for giving up its right to access the unlicensed band should be compensated by all the other MNOs that can benefit from the reduction of channel contention. A mutual agreement must be reached by the right-giving-up MNOs and the MNOs that are willing to provide compensations. Let $\cD$ be the set of MNOs that are willing to give up their rights to access the unlicensed band for $\cD \subset {\cal M}$. How to divide the utility between the right-giving-up MNOs and the rest of the MNOs depends on the detailed requirements and the benefit that can be achieved by each MNO. In this paper, \blu{we employ a common adopted setting called {\em transferrable utility} in which the utility obtained by the MNOs in the unlicensed band can be freely transferred between different member MNOs. } We will give a more detailed description about this framework in the next section.

From the previous discussion, the unlicensed band resources that can be accessed by the $k$th link of MNO $i$ is specified by the probability of channel access $\xi_{k, i \backslash {\cal D}}$. Each MNO can then distribute the channel access at each link according to the QoS of the supported types of services. Let $\alpha^{(l)}_{k,i}$ be the portion of the channel access probability that is allocated to support type $l$ service at link $k$ of MNO $i$. We have $\sum_{l\in {\cal Y}} \alpha^{(l)}_{k,i} = \xi_{k, i \backslash {\cal D}}$. We also write $\balpha^{(l)}_i = \langle \alpha^{(l)}_{k,i} \rangle_{k\in {\cal L}_i}$ and $\balpha_i = \langle \balpha^{(l)}_i \rangle_{l\in {\cal Y}}$. We can write the utility obtained by MNO $i$ from supporting type $l$ service at the $k$th link as $\nu^{(l)}_{k,i} (\alpha^{(l)}_{k,i}) = \rho^{(l)}_i \alpha^{(l)}_{k,i} B^{(u)} R_{k, i}$.

We can write the resource allocation problem in unlicensed bands as
\begin{subequations}\label{eq_UnlicensedSlicing_Orig}
\begin{align}
\max\limits_{\balpha_i}& \sum_{k \in {\cal L}_i} \sum\limits_{l\in {\cal Y}} \nu^{(l)}_{k,i} \left( \alpha^{(l)}_{k,i} \right)\label{eq_UnlicensedSlicing_Obj} \\
\;\; {\rm s.t.} &\;\; \sum\limits_{l\in{\cal Y}} \alpha^{(l)}_{k,i} = \xi_{k, i \setminus {\cal D}} \mbox{ and } 
\alpha^{(l)}_{k,i} B^{(u)} R_{k,i} \ge \eta^{(l)}_{i},\;\;\; \forall k\in {\cal L}_i, l\in {\cal Y}. \nonumber  
\end{align}
\end{subequations}

\subsection{Network Slicing over Licensed and Unlicensed Bands}
\label{Subsection_SlicingLicensedUnlicensedBand}
It can be observed that the network slicing decision made by MNOs in licensed and unlicensed bands can be closely related to each other. In particular, if an MNO cannot secure enough spectrum resource in the licensed band, it will become more aggressive in the unlicensed band and will be willing to pay more for the right of other MNOs. Similarly, if the licensed band can offer sufficient resources for some MNOs to support their traffics, these MNOs will be more willing to sell their right in unlicensed band.

The main objective for each MNO is to carefully decide the resource distributed in both licensed and unlicensed bands for each slice. Let $\varpi^{(l)}_{k,i} \left( \alpha^{(l)}_{k,i}, u^{(l)}_{k,i} \right) = \pi^{(l)}_{k,i} \left( u^{(l)}_{k,i} \right) + \nu^{(l)}_{k,i} \left( \alpha^{(l)}_{k,i} \right)$. Each MNO $i$ decides the optimal resource distribution by solving the following problem:
%
\blu{
\begin{subequations} \label{eq_LicenseUnlicenseOrig}
\begin{align}
\max\limits_{\bu_i, \balpha_i}& \sum_{k \in {\cal L}_i} \sum\limits_{l\in {\cal Y}} \varpi^{(l)}_{k,i} \left( \alpha^{(l)}_{k,i}, u^{(l)}_{k,i} \right) \label{eq_LicenseUnlicenseObj} \\
\;\; {\rm s.t.} &\;\; \sum\limits_{l\in{\cal Y}} \alpha^{(l)}_{k,i} = \xi_{k, i \setminus {\cal D}} \mbox{ and } \sum\limits_{l\in{\cal Y}} u^{(l)}_i \le \sum\limits_{j \in {\cal C}^{(l)}} B_{j}, \label{eq_LicenseUnlicenseCon1}\\
&\;\; (u^{(l)}_{k,i} + \alpha^{(l)}_{k,i} B^{(u)} ) R_{k,i}  \ge \eta^{(l)}_i, \;\;\; \forall k\in {\cal L}_i, l\in {\cal Y} \label{eq_LicenseUnlicenseCon3}
\end{align}
\end{subequations}
where the first term in  (\ref{eq_LicenseUnlicenseCon1}) means that the channel access probability allocated by each link $k$ to support each type $l$ of service cannot exceed the total probability of channel access that can be obtained by each BS. (\ref{eq_LicenseUnlicenseCon3}) means that the combined throughput obtained in both licensed and unlicensed band for each type of services must satisfy the minimum throughout requirement.
}

\section{Network Slicing Game}
\label{Section_Game}
To model the negotiation and interaction among multiple MNOs, we apply the framework of the overlapping coalition formation game. The overlapping coalition formation game attracts much attention recently due to its capability to investigate the resource allocation problem between multiple players that can allocate different portions of their resources to simultaneously support different types of services by joining as members of different coalitions\cite{Chalkiadakis2010OverlapCoalitioanGame}. Compared to the traditional partition-based coalition formation game, allowing players to interaction with each other across multiple coalitions has the potential to further improve the resource utilization efficiency and increase the outcome for the players.
We formally define network slicing game as follows.

\begin{definition}
\label{Definition_OverlapCoalitionForm}
A \emph{network slicing game} is defined by a tuple ${\cal A} = \langle {\cal M}, {\bB}, {\cal Y}, \bvarpi \rangle$ where ${\cal M}$ is the set of MNOs that may share spectrum with each other, $\bB = \cup_{i\in {\cal M}} B_i \times B^{(u)}$ is the spectrum that can be accessed by MNOs in both licensed and unlicensed bands, $\cal Y$ is the set of service types supported by each MNO, 
$\bvarpi$ is the vector of utilities obtained by the MNOs.
\end{definition}

We give a more detailed discussion for each of the above elements in the network slicing game as follows: \blu{each MNO can access licensed and unlicensed band spectrum. The licensed band that can be accessed by each MNO includes both its own licensed band as well as licensed bands owned by other MNOs that can be accessed through inter-operator carrier aggregation described in Section \ref{Subsection_SlicingLicensedBand}. Each MNO can access the unlicensed spectrum through channel contention following the CSMA mechanism. The main objective for each MNO is to slice the shared licensed spectrum as well as the access probability of unlicensed spectrum  to support all types of service. Each type $l$ of service is specified by a threshold $\eta^{(l)}_i$ characterizing the minimum QoS that needs to be be guaranteed by each MNO $i$ and a price $\rho^{(l)}_i$ describing the unit price charged by MNOs for supporting the service}.
A slice $\bc^{(l)}$ is a vector ${\bc^{(l)}} = \langle c^{(l)}_{1}, c^{(l)}_2, \ldots, c^{(l)}_{M} \rangle$ where $c^{(l)}_{i} = \langle w^{(l)}_i, \balpha^{(l)}_i \rangle$ is the resource allocated by MNO $i$ to support type $l$ service. Each slice comprises of portions of licensed spectrum and portions of channel access probability in unlicensed band allocated to the supported service type. The licensed spectrum distributed to support type $l$ service is given by $\bw^{(l)} = \langle w^{(l)}_i \rangle_{i\in {\cal C}^{(l)}}$. 
The access probabilities of the unlicensed band allocated by MNOs to support type $l$ service can be written as $\balpha^{(l)} = \langle \alpha^{(l)}_{k,i} \rangle_{k\in {\cal L}_i, i\in {\cal M}}$. $\alpha^{(l)}_{k,i}=0$ means that MNO $i$ does not access any unlicensed spectrum to support type $l$ service for link $k$.

We define a network slicing structure ${\bc} = \langle \bc^{(l)} \rangle_{l\in {\cal Y}}$ as a vector specifying the resource allocations of all the MNOs among all types of service.

We consider a game with transferrable utility in which the utility obtained in a slice can be freely transferred among contributing MNOs. A characterization function maps each slice of MNOs into a single value referred to as the worth of a slice. The worth characterizes the total utility that is available to all the contributing MNOs. The worth for each slice consists of utilities obtained from both licensed and unlicensed bands. We can write the worth of a slice $\bc^{(l)}$ as
$
v\left(\bc^{(l)}\right) = \sum\limits_{i\in {\rm supp} ({\bc^{(l)}})} \sum\limits_{k\in {\cal L}_i} \varpi^{(l)}_{k,i}$
where ${\rm supp}\left( \cdot \right)$ is the support. 

We can observe that the worth function is monotone. In particular, suppose $\bc^{(l)}$ and  $\bc'^{(l)}$ are two possible slices for supporting type $l$ service. We then have $v\left(\bc^{(l)}\right) \ge v\left(\bc'^{(l)}\right)$ for any $\bc^{(l)}$, $\bc'^{(l)}$ such that $c^{(l)}_i \ge c'^{(l)}_i$ for all $i\in {\cal M}$. In other words, MNOs will always allocate all the accessible spectrum to serve the supported service.

We define an allocation of utility for each slice as $\bx^{(l)} = \langle x^{(l)}_{i} \rangle_{i \in {\rm supp} (\bc^{(l)})}$ which describes the worth distributed among all the MNOs. $\bx^{(l)}$ is efficient if $\sum_{i\in {\rm supp} (\bc^{(l)})} x^{(l)}_i = v\left(\bc^{(l)}\right)$. $\bx^{(l)}$ is also called imputation if it is efficient and satisfies the individual rationality, i.e., $x^{(l)}_i \ge v \left( \bar{c}^{(l)}_{i} \right)$ where $\bar{c}^{(l)}_i$ is the slice for type $l$ service if MNO $i$ cannot cooperate and share spectrum resources with other MNOs.
We define a network slicing agreement as a tuple $\langle \bc, \bx \rangle$ where $\bx = \langle \bx^{(l)} \rangle_{l\in {\cal Y}}$.

As mentioned earlier, MNOs are self-interest entities and always seek to maximize their individual utilities by forming coalitions with different MNOs in both licensed and unlicensed bands. However, the resource distribution and negotiation among MNOs across different slices can be very complex. For example, when an MNO negotiating with another MNO for sharing their resources to serve a specific type of service, it can also offer a certain term that may affect the cooperation with other MNOs in serving other types of service. Similarly, when an MNO deviates from a network slicing agreement with another MNO in serving a specific type of service, it can also affect its cooperation with other MNOs supporting other types of service. The main solution concept in the network slicing game is the {\em core}.
We extend the concept of the conservative core in the overlapping coalition formation game into our network slicing game. 
\begin{definition}
Given a network slicing game ${\cal A} = \langle {\cal M}, \bB, {\cal Y}, {\bvarpi} \rangle$ and a subset of MNOs ${\cal N} \subseteq {\cal M}$.  Suppose $\langle \bc, \boldsymbol{x} \rangle$ and $\langle \bc', \boldsymbol{x}' \rangle$ are two network slicing agreements such that for any slice $\bc^{(l)} \in \bc$ either ${\rm supp} (\bc^{(l)}) \subseteq {\cal N}$ or ${\rm supp} (\bc^{(l)}) \subseteq {\cal M}\setminus {\cal N}$. We say that network slicing agreement $\langle \bc', \boldsymbol{x}' \rangle$ is a profitable deviation of ${\cal N}$ from $\langle \bc, \boldsymbol{x} \rangle$ if for all $j \in {\cal N}$, we have $\varpi_{j} \left( \bc', \boldsymbol{x}' \right) > \varpi_{j} \left( \bc, \boldsymbol{x} \right)$. We say that a network slicing agreement $\langle \bc, \boldsymbol{x} \rangle$ is in the core of $\cal A$ if no subset of $\cal N$ has a profitable deviation from it. In other words, for any subsets of MNOs ${\cal N} \subseteq {\cal M}$, any network slicing structure $\bc_{\cal N}$, and any imputation $\bx'$, we have $\varpi_{j} \left( \bc', \boldsymbol{x}' \right) \le \varpi_{j} \left( \bc, \boldsymbol{x} \right)$.
\end{definition}

We have the following result. 
\begin{theorem}
\label{Theorem_Optimal}
The core of the network slicing game is non-empty and any outcome in the core maximizes the social welfare.
\end{theorem}
\begin{IEEEproof}
See Appendix \ref{Proof_Optimal}.
\end{IEEEproof}

\section{Distributed Optimization Algorithm}
\label{Section_Algorithm}

One of the main challenges for the inter-operator network slicing is to minimize the communication/coordination overhead between MNOs.
In this section, we propose a simple and distributed algorithm framework that can achieve the stable and optimal network slicing structure that is in the core of the network slicing game. Our algorithm is based on the distributed ADMM \cite{Boyd2011ADMM} algorithm to decompose the optimization problem into a set of subproblems. Unfortunately, it is known that the traditional ADMM method can only solve problems consisting of two blocks of random variables and therefore cannot be directly applied to solve problem (\ref{eq_LicenseUnlicenseOrig}) which consists of a large number of variables. In addition, the original ADMM method is a centralized approach that requires all players to reveal their private information to a central controller. Most existing distributed ADMM methods focused on designing a consensus mechanism in which the neighboring agents can exchange and jointly update a local copy of their model parameters\cite{Wei2013ADMM}. These methods cannot be directly applied to solve network slicing problem in inter-operator systems because MNOs are generally unwilling to share their private proprietary information with each other.  

We propose a D-ADMM-PVS algorithm to optimize the network slicing for inter-operator systems. In our algorithm framework, the inter-operator network slicing problem is first divided into $\sum_{i\in {\cal M}}|{\cal L}_i|$ number of sub-problems each of which can be solved by an individual link (can be either UE or BS of the corresponding link) of an MNO using its local information. Each link will submit a single dual variable to its associated MNO and all the MNO will only coordinate their collected dual variable using a linear function.

As observed in Section \ref{Section_Game}, the property of transferrable utility makes MNOs have the incentive to jointly slice their resources and maximize the total social welfare. Let us write the social welfare maximization problem for the network slicing game as follows:
\blu{
\begin{subequations}\label{Orig}
\begin{align}
\max\limits_{\bu, \balpha}& \sum\limits_{i \in {\cal M}} \sum_{k \in {\cal L}_i} \sum\limits_{l\in {\cal Y}} \varpi^{(l)}_{k,i} \left( {u^{(l)}_{k,i}}, \alpha^{(l)}_{k,i} \right) \label{Obj} \\
\;\; {\rm s.t.} &\;\; \sum\limits_{l\in{\cal Y}} \alpha^{(l)}_{k,i} = \xi_{k, i \setminus {\cal D}},  \label{Con1}\\
&\;\; (\bu_{k,i} + \balpha_{k,i} B^{(u)} ) R_{k,i}  \succeq \bdeta_i,  \label{Con3}\\
&\;\; \sum\limits_{l\in{\cal Y}} u^{(l)}_{k,i} \le \sum\limits_{j\in {\cal C}^{(l)}} B_{j}, 0 \le \alpha^{(l)}_{k,i} \le 1, \mbox{ and } u^{(l)}_{k,i} \ge 0, \nonumber \\
&\;\;\;\;\;\;\;\;\;\;\;\;\;  \forall k\in {\cal L}_i, l\in {\cal Y},
\label{Con4}
\end{align}
\end{subequations}
where $\balpha = \langle \balpha_{k,j} \rangle_{k\in {\cal L}_j, j\in {\cal M}}$ and $\balpha_{k,i}=\langle \alpha_{k,i}^{(l)} \rangle_{l\in{\cal Y}}$ are the vector of channel access probability allocated to all supported types of services as well as each link, and $\bu = \langle \bu_j \rangle_{j\in {\cal M}}$, $\bu_{k,i}=\langle u_{k,i}^{(l)} \rangle_{l\in{\cal Y}}$, $\bdeta_{i}=\langle \eta^{(l)}_i \rangle_{l\in{\cal Y}}$ is vector of the minimum QoS required by all the supported types of service offered by MNO $i$, and $\succeq$ is the vector inequality. Note that the constraint in (\ref{Con1}) means that the total probability of channel access as well as the licensed bandwidth allocated for supporting all types of services cannot exceed the available channel access probability  as well as the licensed band owned by each MNO. (\ref{Con4}) means that the aggregate throughput obtained in both licensed and unlicensed band must satisfy the minimum throughput requirement for each type of supported service.
}

Note that in problem (\ref{Orig}), we replace ${\rm supp} ({\bc}^{(l)})$ with the set of all the MNOs $\cal M$. This does not impact our results because the utility division among MNOs in each slice has already been decided by $d^{(l)}_{k,i}$. In other words, even if, due to the limit of the resources, some MNOs choose to not distribute any licensed resource to support a certain type of services (e.g., type $l$ service), this does not mean these MNOs cannot receive benefit from serving type $l$ services for its UEs because they can still access the spectrum resource $\bw^{(l)}$ distributed by other MNOs. In other words, $w^{(l)}_{i} = 0$ does not mean $\pi^{(l)}_{k,i} = 0$.

Let $f_{k,i}(\balpha_{k,i})= \sum_{l\in {\cal Y}} B^{(u)} \rho^{(l)}_i R_{k, i}  \alpha^{(l)}_{k,i}$, and $g(\bu)=\sum_{i \in {\cal M}} \sum_{k \in {\cal L}_i} \left( \sum_{l\in {\cal Y}} \rho^{(l)}_i R_{k, i} \left( \sum_{j \in {\cal M}} u^{(l)}_{k,j} \right) \right)$. We can rewrite the objective function in (\ref{Obj}) as the summation of  a set of sub-functions as follows
\begin{equation}
\sum_{i \in {\cal M}} \sum_{k \in {\cal L}_i} f_{k,i}(\balpha_{k,i}) + g(\bu).
\end{equation}

Let us introduce a set of indicator functions to incorporate constraints (\ref{Con1}) into the objective function. For the separable constraints (\ref{Con1}), we have
\begin{equation}
\cI^{\alpha}_{k,i}(\balpha_{k,i}) =
  \begin{cases}
    0,       & \balpha_{k,i} \in \cE^{\alpha}_{k,i},    \\
    \infty, & \balpha_{k,i} \notin \cE^{\alpha}_{k,i},
  \end{cases},
  \cI^{u}(\bu) =
  \begin{cases}
    0,       & \bu \in \cE^{u},    \\
    \infty, & \bu \notin \cE^{u},
  \end{cases}
\end{equation}
where
\begin{eqnarray}
\cE^{\alpha}_{k,i}=\{\balpha_{k,i}|\sum_{l\in {\cal Y}} \alpha^{(l)}_{k,i} = \xi_{k, i \setminus {\cal D}},0 \le \alpha^{(l)}_{k,i} \le 1\},
\label{eq_Indicatorconstraint_alpha}\\
\cE^u=\{\bu|\sum_{l\in {\cal Y}} u^{(l)}_{k,i} \leq B_i, u^{(l)}_{k,i} \ge 0, \forall i \in \cM\},
\label{eq_Indicatorconstraint_u}
\end{eqnarray}
\blu{where (\ref{eq_Indicatorconstraint_alpha}) and (\ref{eq_Indicatorconstraint_u}) correspond to the ranges of constraints (\ref{Con1}) and (\ref{Con4}), respectively. }
We can incorporate them into objective functions as follows:
\begin{eqnarray}
f^+_{k,i}(\balpha_{k,i})=f_{k,i}(\balpha_{k,i} ) + \cI^{\alpha}_{k,i}, \; g^+(\bu)=g(\bu)+\cI^{u}(\bu).
\label{eq_fg}
\end{eqnarray}
\blu{where the first and second terms in (\ref{eq_fg}) correspond to the objective function in (\ref{Obj}) incorporated with the constraints (\ref{Con1}) and (\ref{Con4}). }

For the inseparable constraint (\ref{Con3}), we also introduce an indicator function as follows
\begin{equation}
\cI_Z(\bX) =
  \begin{cases}
    0,       & \bX \in \cE_Z,    \\
    \infty, & \bX \notin \cE_Z,
  \end{cases},
\end{equation}
where $\bX=(\balpha, \bw)$, and
\begin{equation}
\cE_Z = \{\bX|( \bu_{k,i} + \balpha_{k,i} B^{(u)} ) R_{k,i}  \succeq \bdeta_i\}.
\end{equation}

We can then reformulate the above optimization problem in ({\ref{Orig}}) as
\begin{subequations}\label{Reform}
\begin{align}
 \max_{\bX,\bZ} & \quad \quad F(\bX)+ \cI_Z(\bZ), \label{Obj_R} \\
\mbox{s.t.}  &\quad \quad \bX-\bZ=\bzero, \label{Con_R}
\end{align}
\end{subequations}
where $F(\bX)=\sum_{i \in {\cal M}} \sum_{k \in {\cal L}_i} f^+_{k,i}(\balpha_{k,i}) + g^+(\bw)$, and $\bZ$ is the auxiliary variable introduced to isolate the inseparable constraint. The augmented Lagrangian of problem (\ref{Reform}) is
\begin{equation}
\cL_{\gamma}(\bX,\bZ,\bLambda)=F(\bX)+\cI_Z(\bZ)
+\bLambda^T(\bX-\bZ)+\frac{\gamma}{2}\parallel \bX - \bZ \parallel_2^2,
\label{eq_AugLagrangianForm}
\end{equation}
where $\vartheta > 0$ is the augmented Lagrangian parameter, and $\bLambda$ is the dual variable. We can then follow the same line as standard ADMM and write the centralized solution for (\ref{Reform}) as follows:
\begin{subequations}
\begin{align}
\bX^{(t+1)} & := \mbox{arg}\min_{\bX} F(\bX) + \frac{\vartheta}{2}\parallel \bX-\bZ^{(t)} + \bLambda^{(t)} \parallel^2_2, \label{XSub}\\
\bZ^{(t+1)} & := \mbox{arg}\min_{\bZ} \cI_Z(\bZ) + \frac{\vartheta}{2}\parallel \bX^{(t+1)}-\bZ + \bLambda^{(t)} \parallel^2_2, \label{ZSub}\\
\bLambda^{(t+1)} &\;= \bLambda^{(t)}+ \bX^{(t+1)} - \bZ^{(t+1)}, \label{DSub}
\end{align}
\end{subequations}
where we use superscript $(t)$ to denote the $t$th iteration.

To solve the above problem in a distributed manner, we split variable $\bX$ into a set of sub-vectors, namely $\balpha_{k,i}$ and $\bw_i$. We also separate the $\bX$-updating step into a set of sub-problems as follows.

Each link $k$ solves the following $\balpha_{k,i}$-subproblem for unlicensed band resource distribution using its local information:
\begin{equation}\label{eq_LinkOptmSubProb} 
\balpha_{k,i}^{(t+1)}= \mbox{arg}\min_{\balpha_{k,i}}  \;\; f^+_{k,i}(\balpha_{k,i}) + \frac{\vartheta}{2}\parallel \balpha_{k,i}-\bZ_{k,i}^{(t)} + \bLambda_{k,i}^{(t)} \parallel^2_2;
\end{equation}

MNOs will jointly solve the following $\bw$-subproblem for licensed band resource distribution through DSM function block in the 3GPP network sharing framework:
\begin{equation}\label{w_sub}
\bw^{(t+1)}= \mbox{arg}\min_{\bw}  \;\; g^+(\bw) + \frac{\vartheta}{2}\parallel \bw
-\bZ_{w}^{(t)} + \bLambda_{w}^{(t)} \parallel^2_2;
\end{equation}

A coordinator deployed in the DSM block is responsible for coordinating the $\bZ$-updating in (\ref{ZSub}) and the $\bLambda$-updating in (\ref{DSub}). 
%
We summarize the details of the proposed algorithm in Algorithm~\ref{Algorithm_ADMM}. The convergence rate is presented in Theorem~\ref{Theorem_ADMM}.

\begin{algorithm}
  \caption{D-ADMM-PVS Algorithm}\label{Algorithm_ADMM}
  \begin{algorithmic}

    \STATE \textit{Initialization}: $\balpha ^0 $, $\bw^0$, $\gamma>0$, t=1;
    \FOR {$t=1,2,...$}
    \STATE 1. Each UE or BS corresponding to the $k$th link executes the following steps:
    \STATE \quad 1a) Update $\balpha_{k,i}^{(t+1)}$ according to (\ref{eq_LinkOptmSubProb});
    \STATE \quad 1b) Report $\balpha_{k,i}^{(t+1)}$ to the corresponding MNO $i$;
    \STATE 2. MNOs collect the intermediate results $\balpha_{k,i}^{(t+1)}$ from their UEs and/or BSs, and report them to the SDN orchestrator.
    \STATE 3. The regional orchestrator executes the following steps:
    \STATE \quad 3a) Sequentially update $\bw$, $\bZ$ and $\bLambda$ by following (\ref{w_sub}),(\ref{ZSub}) and (\ref{DSub});
    \STATE \quad 3b) Feedback the auxiliary variables $\bZ_{\cdot,i}$ and the dual variables $\bLambda_{\cdot,i}$ to the corresponding MNO $i$;
    \STATE 4. MNO $i$ feedbacks auxiliary variable $\bZ_{k,i}$ and dual variable $\bLambda_{k,i}$ to the UE or BS associated with link $k$.
    \IF{Stopping criteria meets}
    \STATE break;
    \ENDIF
    \STATE $t=t+1$
    \ENDFOR
  \end{algorithmic}
\end{algorithm}

In the above algorithm, each MNO $i$ needs to solve problem (\ref{eq_LinkOptmSubProb}) using its own private information related to each of its links. Each MNO $i$ sends its calculated solution $\bZ_{\cdot, i}$ to the local regional orchestrator. The local orchestrator will then update the dual variable $\bLambda_{\cdot,i}$ collected from all the MNOs and send each MNO with its individual dual variable related to the sub-problem in (\ref{eq_LinkOptmSubProb}). In other words, in D-ADMM-PVS Algorithm, each MNO does not need to disclose its private information and still can achieve the global optimal solution of problem  (\ref{Orig}).

We can prove the following result.
\begin{theorem}\label{Theorem_ADMM}
The augmented Lagrangian form of the objective function for problem (\ref{Obj_R}) is separable and convex. Algorithm 1 maximizes the social welfare.
\end{theorem}
\begin{IEEEproof}
See Appendix \ref{Proof_ADMM}.
\end{IEEEproof}

\section{Performance Evaluation}

\begin{figure}
\begin{minipage}[t]{1\linewidth}
\centering
\includegraphics[width=2.0 in]{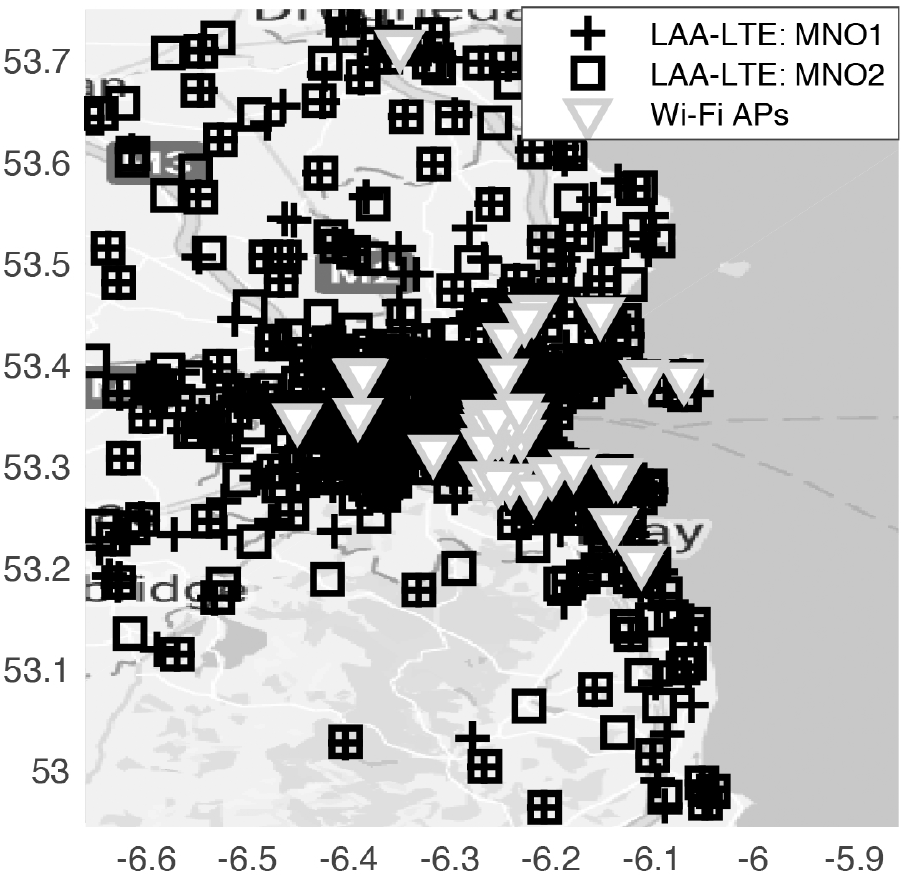}
\vspace{-0.1in}
\caption{Distribution of BSs deployed by two major MNOs as well as Wi-Fi hotspots deployed at Starbucks in the city of Dublin.}
\label{Figure_BSlocations}
\end{minipage}
%
\begin{minipage}[t]{1\linewidth}
\centering
\tiny
\begin{tabular}{|l|l|}
\hline
Parameter & Value \\
\hline \hline
Wi-Fi traffic class & Voice (AC = VO) \\
\hline
LAA traffic class & Voice (PC = 1) \\
\hline
PHY rate& $52$ Mbps\\
\hline
Unlicensed bandwidth& $20$  MHz\\
\hline
Transmission power& $23$ dBms\\
\hline
LAA noise floor& $-100$ dBm\\
\hline
Wi-Fi noise floor& $-90$ dBm\\
\hline
Path Loss Model& $43.3\log(d) + 11.5 + 20\log(f_c)$\\
\hline
Wi-Fi CCA threshold& $-62$ dBm\\
\hline
LAA CCA threshold& $-62$ dBm\\
\hline
\end{tabular}
\caption{Simulation Parameters}
\label{tb:sim_parameters}
\end{minipage}
%
\begin{minipage}[t]{1\linewidth}
\centering
\includegraphics[width=2.0 in]{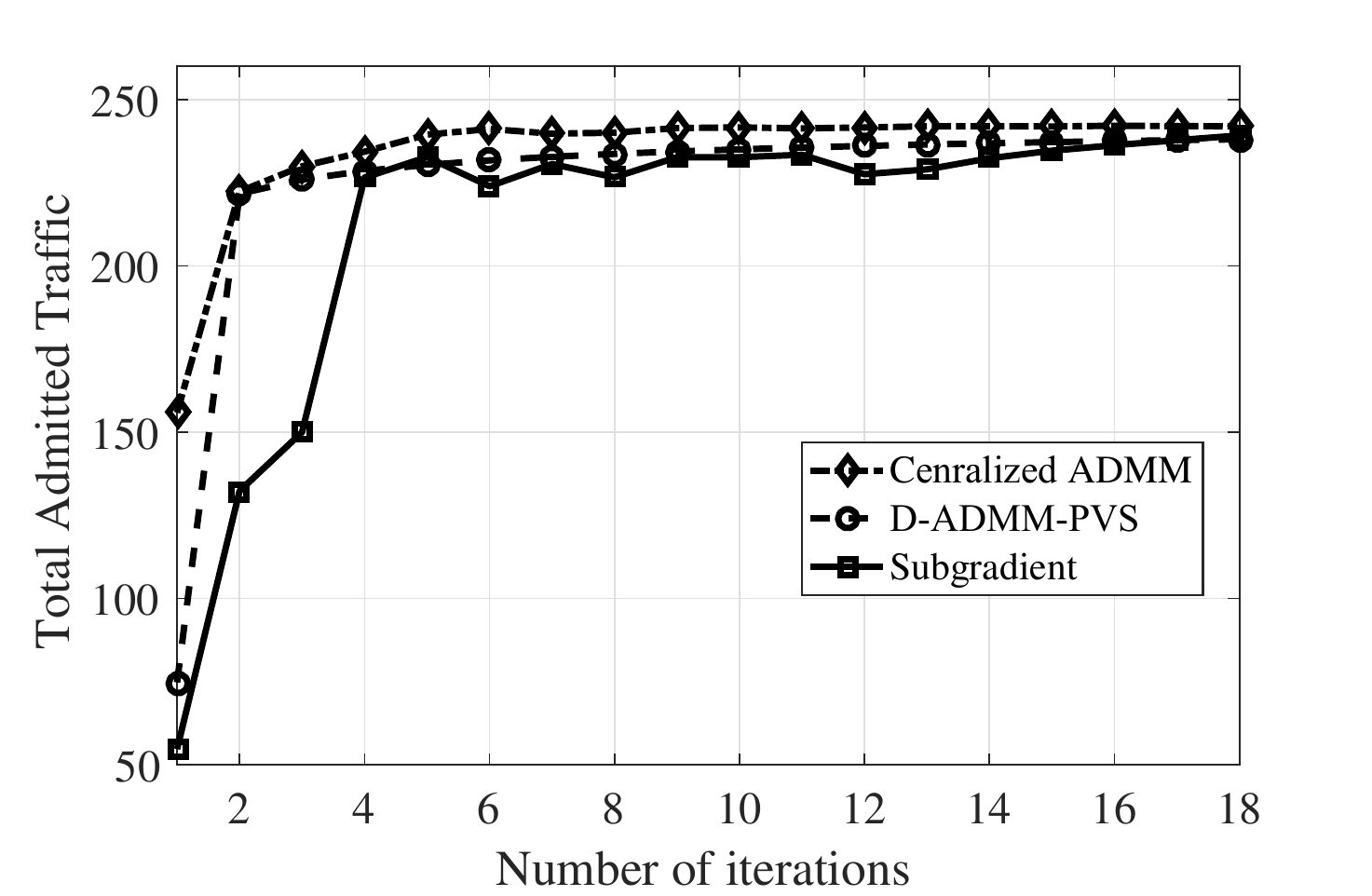}
\vspace{-0.1 in}
\caption{Convergence rate of D-ADMM-PVS Algorithm compared to subgradient and centralized ADMM algorithms.}
\label{Figure_Converge}
\end{minipage}
\end{figure}

\begin{figure}
\begin{minipage}[t]{0.48\linewidth}
\centering
\includegraphics[width=1.8 in]{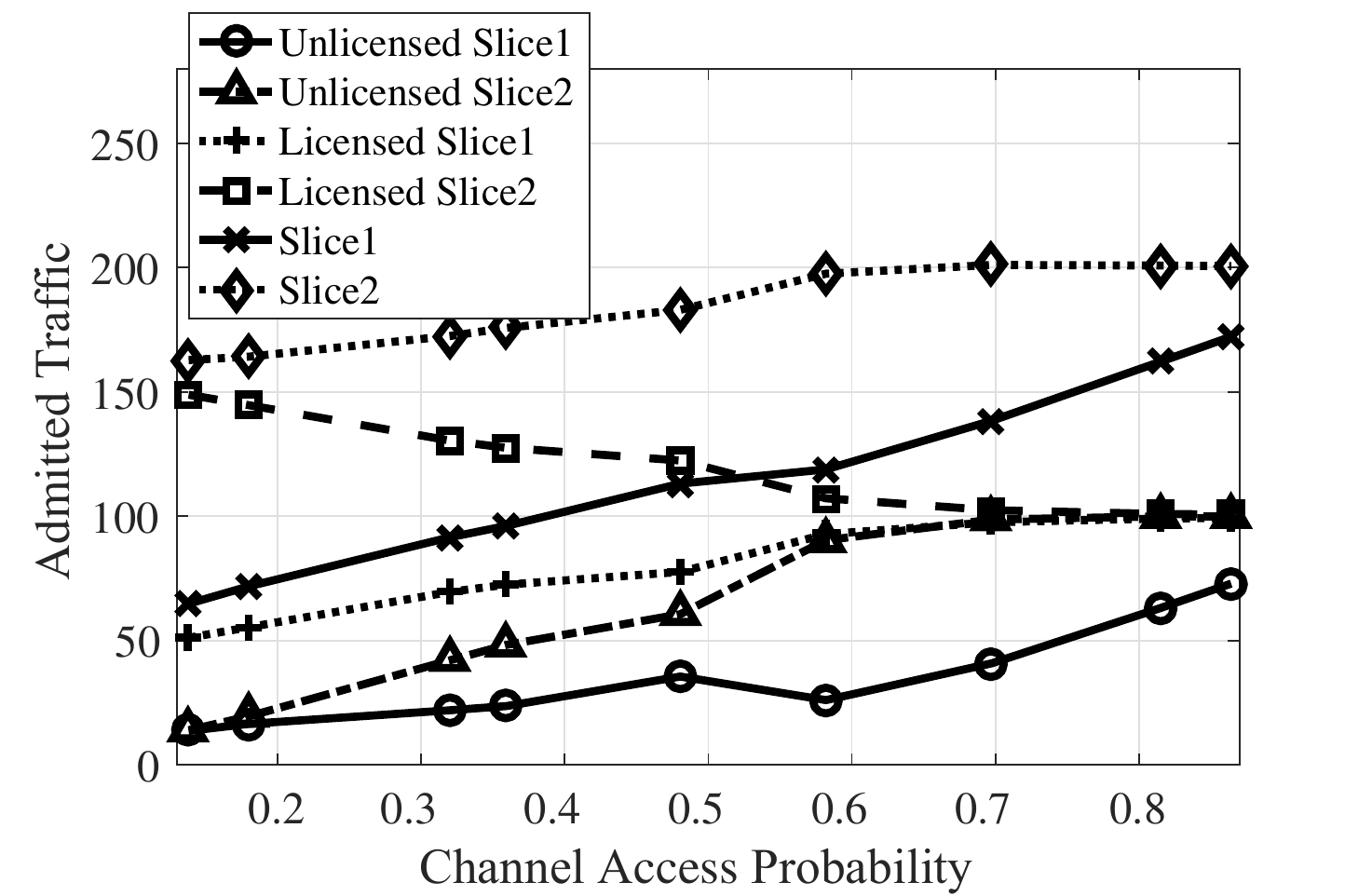}
\vspace{-0.3in}
\caption{Traffic admitted by each slice under different network densities.}
\label{Figure_Xi1}
\end{minipage}
%
\begin{minipage}[t]{0.5\linewidth}
\centering
\includegraphics[width=1.8 in]{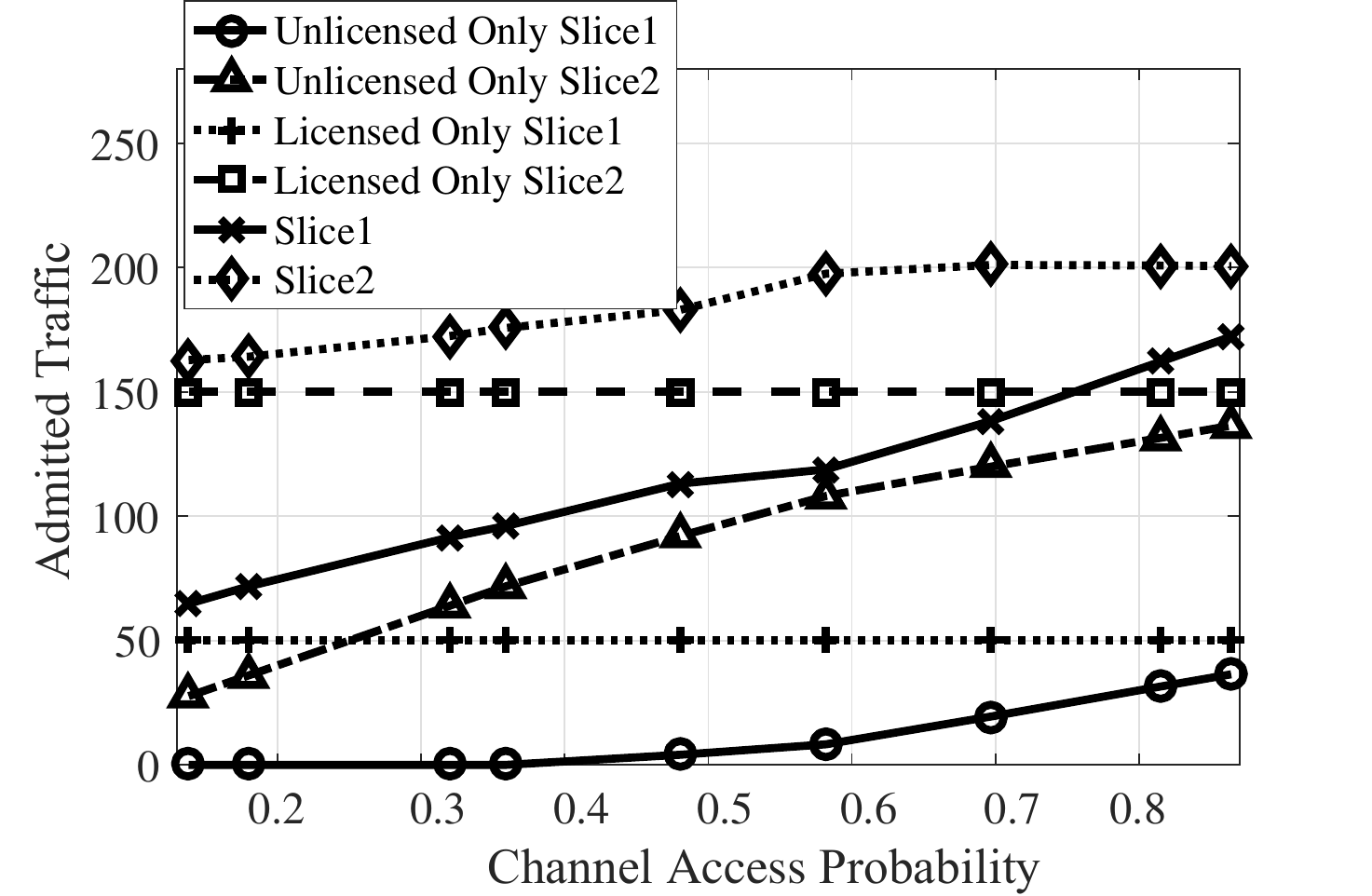}
\vspace{-0.3in}
\caption{Traffic admitted by network slicing under different network densities.}
\label{Figure_Xi2}
\end{minipage}
%
%
\begin{minipage}[t]{0.48\linewidth}
\centering
\includegraphics[width=1.8 in]{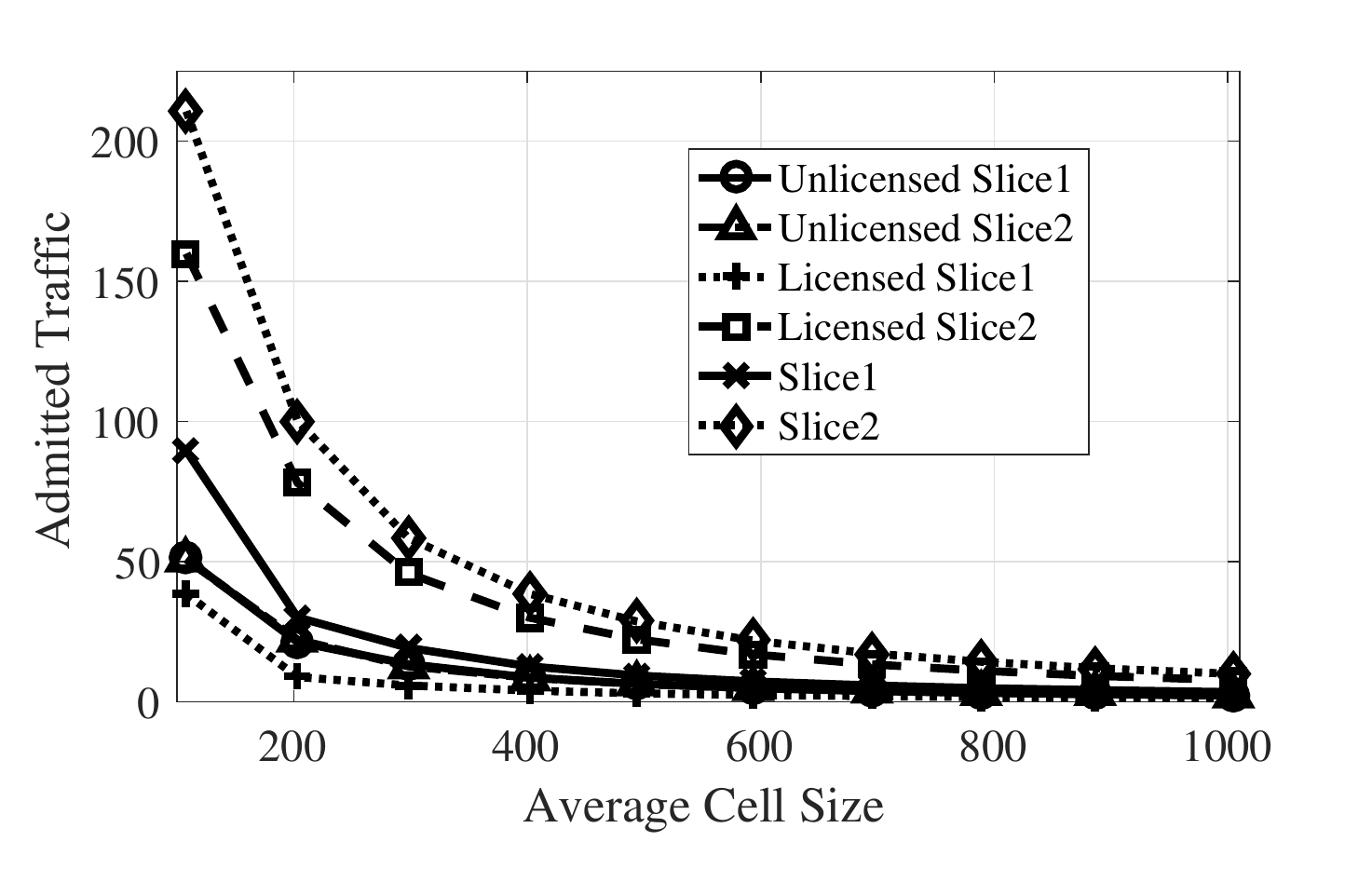}
\vspace{-0.3in}
\caption{Traffic admitted by each slice under different cell sizes.}
\label{Figure_R1}
\end{minipage}
%
\begin{minipage}[t]{0.5\linewidth}
\centering
\includegraphics[width=1.8 in]{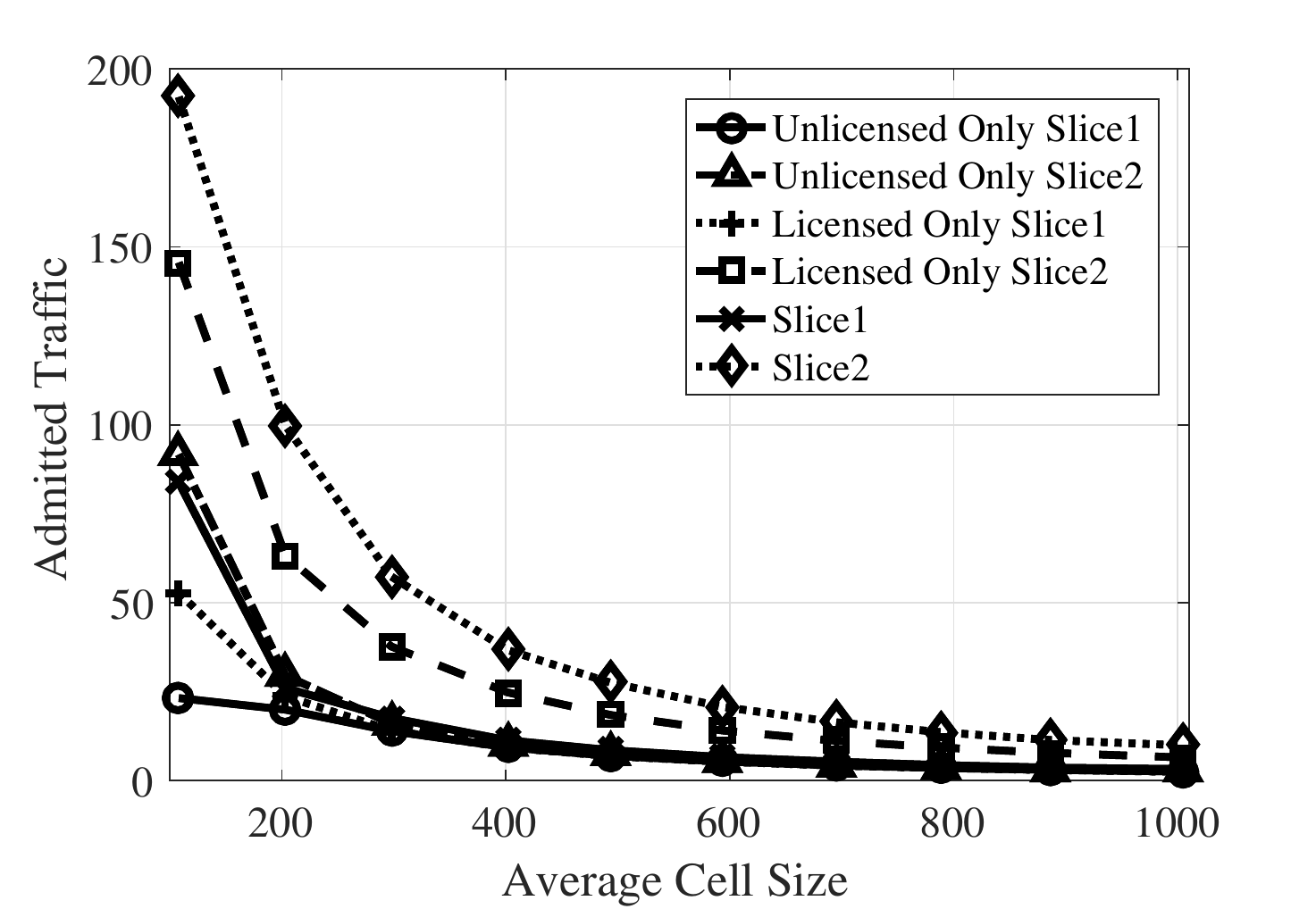}
\vspace{-0.3in}
\caption{Traffic admitted by network slicing band under different cell sizes.}
\label{Figure_R2}
\end{minipage}\\
%
\begin{minipage}[t]{0.48\linewidth}
\centering
\includegraphics[width=1.8 in]{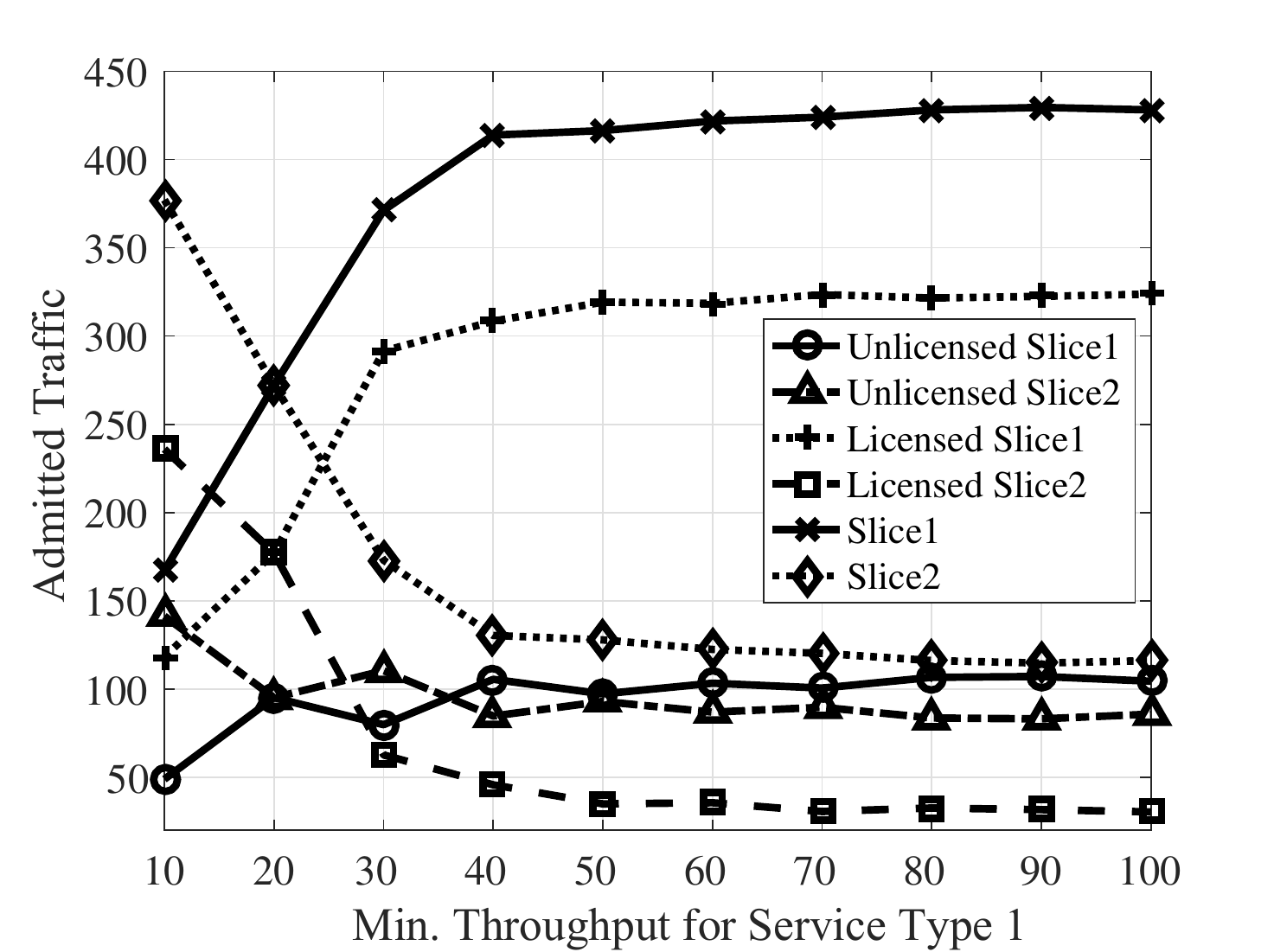}
\vspace{-0.3in}
\caption{Traffic admitted by each slice under different min QoS guarantee.}
\label{Figure_eta1}
\end{minipage}
%
\begin{minipage}[t]{0.5\linewidth}
\centering
\includegraphics[width=1.8 in]{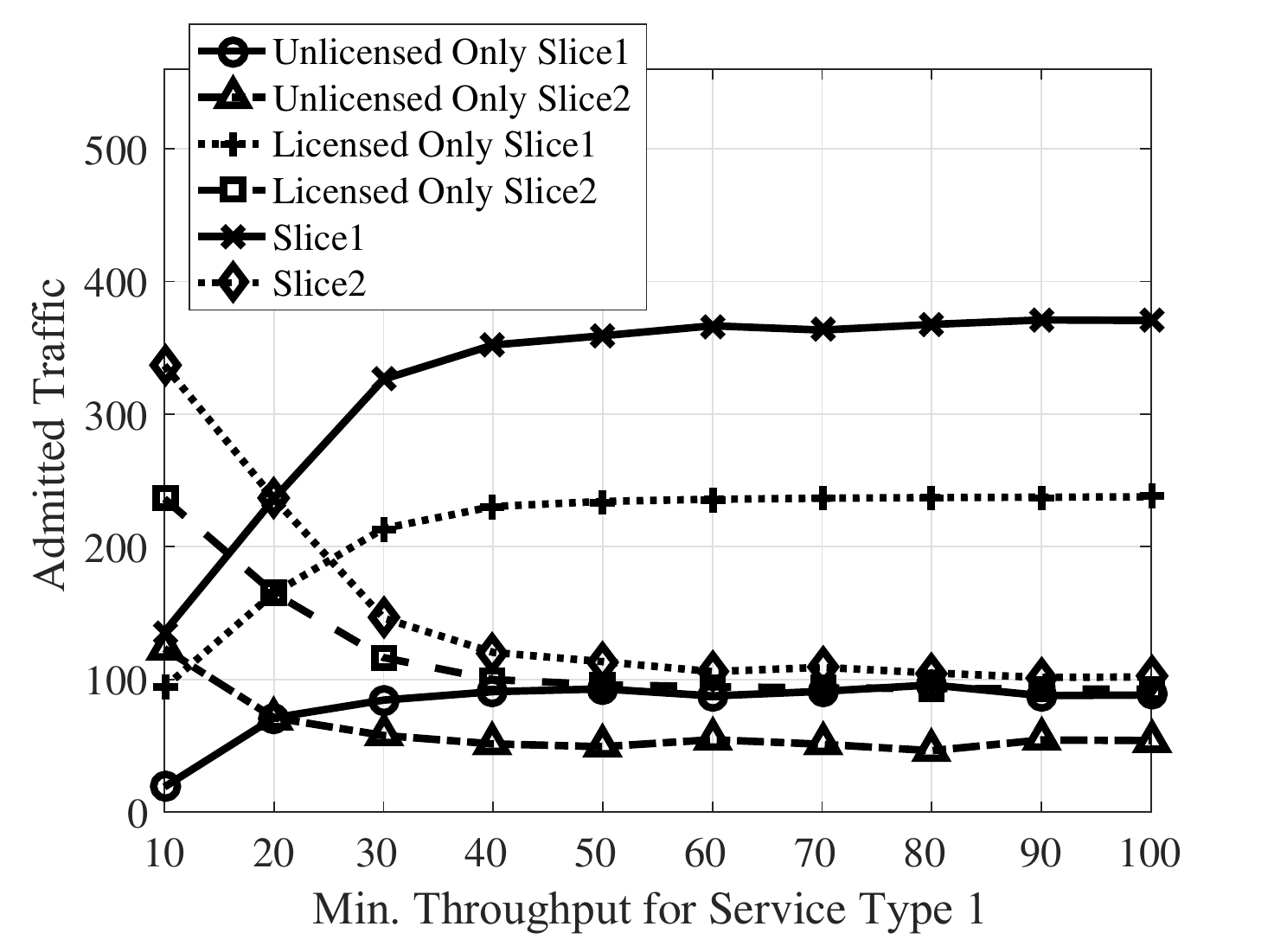}
\vspace{-0.3in}
\caption{Traffic admitted by network slicing under different min QoS guarantee.}
\vspace{-0.5in}
\label{Figure_eta2}
\end{minipage}
\vspace{-0.2in}
\end{figure}

\subsection{Simulation Setup}
To evaluate the performance of our proposed inter-operator network slicing, \blu{we simulate a multi-operator network slicing architecture using over 400 real BS locations (including both GSM and UMTS BSs) in the city of Dublin deployed by two major telecom operators in Ireland\cite{Kibilda2016BSModel}. We assume each BS can operate in both licensed and unlicensed bands. These BSs are coexisting with the Wi-Fi APs installed at 54 Starbucks coffee shops throughout the city\cite{Starbucks}. We focus on the downlink communication from BSs and Wi-Fi APs.  The actual locations and deployment densities of LAA BSs and Wi-Fi APs are presented in Figure \ref{Figure_BSlocations}. We consider the Poisson frame arrival rate with intensity $\lambda=1000$ frames per second for all traffics generated by BSs and Wi-Fi APs. Each MNO can allocate at most 20MHz bandwidth of its licensed band and/or negotiate and trade their access right for a 20MHz unlicensed band in 5.5 GHz unlicensed band. 
}

To evaluate the channel access of coexisting LAA BSs and Wi-Fi APs in unlicensed band, we develop a C++-based discrete event simulator using the CSIM development toolkit\cite{CSIM_lib} with total 3000+ lines of codes to simulate the scheduling and contention behavior 
between LAA BSs as well as that between LAA BSs and Wi-Fi APs.

We implement the most recent LTE-LAA\cite{3GPP2016LAA} and IEEE 802.11ac standards\cite{WiFi80211ac} with channel access parameters listed in Figure \ref{tb:sim_parameters}. We consider the scenario that both LTE BS and Wi-Fi APs adopt the contention parameters according to the traffic class voice (PC=1 for LTE-LAA and AC=VO in Wi-Fi). BSs and Wi-Fi APs have the same transmit power 23 dBm and adopt the same CCA threshold -62 dBm.  We conduct each simulation for $10$ seconds and collect all the traces and logs from all BSs and Wi-Fi APs, including time stamps for frame arrival to MAC queue, time spent in queue, time spent during contention, time spent during transmission.
We consider two types of service traffics (e.g., audio and video streaming) supported by both MNOs requiring 10 Mbps and 20 Mbps minimum guaranteed throughput, respectively.

\subsection{Numerical Results}
In Figure \ref{Figure_Converge}, we compare the convergence performance of our proposed D-ADMM-PVS algorithm with the subgradient with dual decomposition algorithm, one of the most popular distributed optimization algorithms due to its low computation complexity. We can observe that D-ADMM-PVS algorithm converges to the optimal network slicing solution within the first few iterations. We also present the convergence rate when a centralized ADMM in \cite{Boyd2011ADMM} can be implemented to control the network slicing in a centralized fashion. In this case, a centralized controller (can be deployed according to 3GPP shared radio access network architecture) can collect all the information from the MNOs and calculate the spectrum allocation for each slice among all the MNOs. It can be easily observed that this centralized implementation may result in a large communication overhead between MNOs. We can observe that our proposed D-ADMM-PVS algorithm presents a very similar convergence performance as the centralized ADMM and can approach a neighborhood of the optimal solution within the first 2-3 iterations which is much faster than the subgradient method.

It is obvious that the decrease of BS deployment density may result in reduced contention among LAA BSs as well as the increase of channel access for each LAA link in unlicensed band. Therefore, we have carefully chosen nine subregions from the rural areas to the city center of Dublin with different BS deployment densities to evaluate the impact of the network density on the performance of network slicing. We first investigate the total traffic that can be admitted for each type of service when MNOs can cooperate and jointly slice both licensed and unlicensed band. In particular, in Figure \ref{Figure_Xi1}, we present the total traffic admitted in licensed and unlicensed bands by our proposed inter-operator network slicing framework. We observe that allowing MNOs to jointly access licensed and unlicensed bands can significantly increase the traffic volume admitted for all the supported services. Interestingly, we can observe that the portion of the admitted traffic for type 1 service in the licensed band decreases with the channel access probability in unlicensed band. 
This is because the unlicensed band is free and when the channel access probability becomes large, it is more economic for MNOs to offload more traffic from licensed band to unlicensed band.

We then investigate the impact of allowing licensed band spectrum sharing and/or unlicensed band right sharing on the overall traffic that can be admitted for each individual type of service. In Figure \ref{Figure_Xi2}, we consider three different scenarios: S1) {\em unlicensed band only slicing} in which each MNO can only share and trade spectrum access right of unlicensed band with other MNOs, S2) {\em licensed band only slicing} in which MNOs only share and jointly slice the licensed band spectrum, and S3) {\em joint slicing of licensed and unlicensed band}. We observe that the volume of traffic admitted for type 1 service is always more than that admitted for type 2 service. This is because type 2 service has a higher minimum throughput requirement and also charge a higher prices compared to the type 1 service. Therefore, it is more profitable for MNOs to serve more traffic for type 2 service.

It is known that the traffic that can be admitted by each service type is also closely related to the infrastructure deployment density. Generally speaking, the higher the density of BSs, the more traffics can be requested and served by each individual BS. In Figures \ref{Figure_R1}, we assume each cell serves equal number of UEs uniformly randomly located in the coverage area and study the impact of the cell size on the total traffic that can be supported for each type of service. We consider various popular cell sizes from small cell ($\approx$ 100 meter) to macro cell ($\approx$ 1000 meter). We observe that the admitted traffic decreases with the cell size. This is because with the increase of the cell size, the average distance between UEs and the associated BS also increases. This limits the total number of UEs that can be served with the minimum throughput requirement.
We can also observe that
when the cell size becomes large, the unlicensed band become less congested and therefore more traffic will be sent through the unlicensed band for both types of services. Also unlike the licensed band slicing in which the traffic admitted for type 2 service is much higher than that admitted for type 1 service, the traffic transmitted through the unlicensed bands for both types of service are very similar.
In Figure \ref{Figure_R2}, we again consider different scenarios S1)-S3) to compare the performance of MNOs when they can only cooperate in either licensed or unlicensed band, or both. We observe that if MNOs can only cooperate in the unlicensed band, more type 2 service traffic will be offloaded to the unlicensed band compared to the type 1 service.

For a limited network resource, the traffic volume that can be supported for each type of services is also affected by the QoS requirement, that is the minimum throughput that must be guaranteed. The higher the required minimum throughput, the smaller the volume of traffic that can be admitted for the service. To study the impact of the QoS requirement on the traffic that can be supported by the licensed and unlicensed band resources, we fix the minimum throughput required by one particular type of service (service type 2) to $\eta^{(2)}_{i} = 20$ Mbps and compare the traffic volumes admitted by each supported service under different minimum throughput required by the other service type (service type 1) in Figures \ref{Figure_eta1}. We observe that the traffic that can be admitted by  service type 2 decreases with the minimum throughput required by service type 1. In other words, the MNOs tend to obtain more benefit from the service that has a higher QoS requirement. This is consistent with our observation in Figures \ref{Figure_Xi1}--\ref{Figure_R2}. However, the increasing speed of the admitted traffic becomes slower and when the minimum throughput of service type 1 increases above a certain threshold (e.g., 40Mbps), the traffic volume admitted to each supported service approaches to the maximum throughput that can be supported by the available licensed and unlicensed band resources. 

In Figure \ref{Figure_eta2}, we compare the admitted traffic when MNOs can only cooperate in either licensed or unlicensed band, i.e., scenarios S1)-S3), as mentioned at the beginning of this subsection. We observe that both licensed and unlicensed band only scenarios exhibit a similar trend, that is, most of the admitted traffic is associated with the service type that has the higher minimum throughput requirement. We also observe that the total traffic that can be admitted by unlicensed band only slicing is much lower than that admitted by the licensed band only slicing scenario due to the limited channel occupancy time (TXOP duration) for each successful channel access.

\section{Conclusion}
This paper investigates the inter-operator network slicing over licensed and unlicensed bands. 
We develop inter-operator spectrum aggregation method for licensed band slicing and introduce the concept of right sharing for inter-operator network slicing in unlicensed band. 
An mBoE method has been introduced for each MNO to evaluate its benefit in unlicensed band with and without the possible contention from other coexisting MNOs as well as other wireless technologies. 
An overlapping coalition formation game-based framework called network slicing game has been formulated to study the interaction between MNOs in licensed and unlicensed bands.
To reduce the communication overhead between MNOs and preserve the private information of each MNO, we develop a distributed optimization algorithm based on D-ADMM-PVS for inter-operator network slicing.
To evaluate the practical performance of our proposed framework, we consider the possible implementation of LAA BSs over 400 real BS locations of two major MNOs as well as the real Wi-Fi AP locations in the city of Dublin. We develop a C++-based discrete-event simulator to evaluate the contention behavior of LAA BSs and Wi-Fi APs in unlicensed band.
Our numerical results show that our proposed network slicing framework significantly increases the admitted traffics for all supported services especially in the urban environment with high BS deployment density.

\section*{Acknowledgment}
The authors would like to thank Professor Luiz A. DaSilva and Dr. Jacek Kibilda at CONNECT, Trinity College Dublin to provide the BS location data of Dublin.

\appendices
\section{Proof of Theorem \ref{Theorem_Optimal}}
\label{Proof_Optimal}
To prove the core of the network slicing game is always non-empty, we need to first prove that the network slicing game belongs to a special overlapping coalition game that satisfies the property called `convexity'. This means that a coalition can obtain more benefit when it joins a larger coalition. Let ${\cal F} \left( {\cal M} \right)$ be the set of all feasible network slicing agreements. We abuse the notation and use $\bc^{\cal C}$ to denote a network slicing agreement mutually agreed by MNOs in coalition ${\cal C}$. We give a formal definition as follows.
\begin{definition}\cite[Definition 13]{Chalkiadakis2010OverlapCoalitioanGame}
An overlapping coalition formation game is convex if for each ${\cal C} \subseteq {\cal M}$ and ${\cal N} \subset {\cal O} \subseteq {\cal M} \backslash {\cal C}$, the following condition holds: for any ${\bc^{\cal N}, \bx^{\cal N}} \in {\cal F} \left( {\cal N} \right)$, any $\langle \bc^{\cal O}, \bx^{\cal O} \rangle \in {\cal F} \left( {\cal O} \right)$, and any $\langle \bc^{{\cal N}\cup{\cal C}}, \bx^{{\cal N}\cup{\cal C}} \rangle \in {\cal F} \left( {\cal N}\cup{\cal C} \right)$ that satisfies $\bvarpi_{i} \left( \bc^{{\cal N}\cup{\cal C}}, \bx^{{\cal N}\cup{\cal C}} \right) \ge \bvarpi_{i} \left( {\bc^{\cal N}, \bx^{\cal N}} \right)$, $\forall i \in {\cal N}$, there exists an outcome $\langle \bc^{{\cal O}\cup{\cal C}}, \bx^{{\cal O}\cup{\cal C}} \rangle \in {\cal F} \left( {\cal O}\cup{\cal C} \right)$ such that  $\bvarpi_{i} \left( \bc^{{\cal O}\cup{\cal C}}, \bx^{{\cal O}\cup{\cal C}} \right) \ge \bvarpi_{i} \left( {\bc^{\cal O}, \bx^{\cal O}} \right)$, $\forall i \in {\cal O}$ and $\bvarpi_{i} \left( \bc^{{\cal O}\cup{\cal C}}, \bx^{{\cal O}\cup{\cal C}} \right) \ge \bvarpi_{i} \left( \bc^{{\cal N}\cup{\cal C}}, \bx^{{\cal N}\cup{\cal C}} \right)$, $\forall i \in {\cal C}$.
\end{definition}

We have the following lemma.
\begin{lemma}
A network slicing game is convex.
\end{lemma}
\begin{IEEEproof}
We can observe that the utility function of the network slicing game is a linear function of all the possible licensed and unlicensed bands that can be accessed by all the cooperative MNOs. In other words, the more MNOs join the same coalition, the more licensed bands as well as the unlicensed spectrum access rights can be accessed by the member MNOs. We can observe that problem (\ref{eq_LicensedSlicing_Obj}) is a linear function of $w^{(l)}_i$ and $\alpha^{(l)}_{k,i}$. In addition, as mentioned in Sections \ref{Section_Slicing} and \ref{Section_Game}, each MNO will only form a coalition with other MNOs if it cannot obtain a higher utility by forming a coalition with other subsets of MNOs. Let us write the solution of problem (\ref{eq_LicensedSlicing_Obj}) as $\langle \bw^{{\cal C}*}_i, \balpha^{{\cal C}*}_i \rangle$ when the maximum set of MNOs that can share their spectrum with each other to support all services is given by $\cal C$. We can apply the standard convex optimization method to prove that the solution $\varpi_{i}\left( \langle \bw^{{\cal C}*}_i, \balpha^{{\cal C}*}_i \rangle \right)$ satisfies the following properties:
\begin{eqnarray}
\varpi_{i}\left( \langle \bw^{{{\cal O}\cup{\cal C}}*}_i, \balpha^{{{\cal O}\cup{\cal C}}*}_i \rangle \right) &\ge& \varpi_{i}\left( \langle \bw^{{{\cal O}}*}_i, \balpha^{{{\cal O}}*}_i \rangle \right),  \nonumber \\
\varpi_{i}\left( \langle \bw^{{{\cal O}\cup{\cal C}}*}_i, \balpha^{{{\cal O}\cup{\cal C}}*}_i \rangle \right) &\ge& \varpi_{i}\left( \langle \bw^{{{\cal N}\cup{\cal C}}*}_i, \balpha^{{{\cal N}\cup{\cal C}}*}_i \rangle \right), \nonumber \\
&&\;\;\;\;\;\; \forall {\cal N} \subset {\cal O} \subseteq {\cal M} \backslash {\cal C}.
\end{eqnarray}

We can therefore claim that the network slicing game is convex. This concludes the proof.
\end{IEEEproof}

We can then use the following theorem given in \cite{Chalkiadakis2010OverlapCoalitioanGame} to prove the non-emptiness of the core for any network slicing game.
\begin{theorem}\cite[Theorem 3]{Chalkiadakis2010OverlapCoalitioanGame}
If an overlapping coalition formation game is convex, and the worth $v$ is continuous, bounded, monotone and the maximum number of partial coalitions that each MNO can be involved in is finite, then the core of the game is not empty.
\end{theorem}

From Section \ref{Section_Game}, we can directly observe that the worth of the network slicing game satisfies all the above conditions. Therefore, we can claim that the core of the network slicing game is always non-empty. From the definition of the core and following the same line as \cite{Chalkiadakis2010OverlapCoalitioanGame}, we can also prove that a network slicing agreement $\langle \bc, \bx \rangle$ is in the core if and only if
\begin{eqnarray}
\sum_{i\in {\cal M}} \varpi_{i} \left( \bc, \bx \right) \ge v^* \left({\cal M}\right),
\end{eqnarray}
where $v^* \left({\cal M}\right)$ is the supremum of $v \left({\cal M}\right)$. In other words, any outcome in the core maximizes the social welfare. This concludes the proof.

\section{Proof of Theorem \ref{Theorem_ADMM}}
\label{Proof_ADMM}
We can observe that $f_{k,i} \left( \balpha_{k,i} \right)$ is a linear function of $\alpha^{(l)}_{k,i}$ and $g\left( \bw \right)$ is a linear function of $\bw$. Therefore, we can claim that the objective function as well as its corresponding augmented Lagrangian form in (\ref{eq_AugLagrangianForm}) is convex and separable. We can then following the same line as the standard ADMM approach in \cite{Boyd2011ADMM} to prove optimality and the linear convergence rate of Algorithm 1.

\bibliography{reference}
\bibliographystyle{IEEEtran}

\begin{IEEEbiography}{Yong Xiao}(S'09-M'13-SM'15) received his B.S. degree in electrical engineering from China University of Geosciences, Wuhan, China in 2002, M.Sc. degree in telecommunication from Hong Kong University of Science and Technology in 2006, and his Ph. D degree in electrical and electronic engineering from Nanyang Technological University, Singapore in 2012. Currently, he is a professor in the School of Electronic Information and Communications at the Huazhong University of Science and Technology (HUST), Wuhan, China. His research interests include machine learning, game theory, distributed optimization, and their applications in cloud/fog/mobile edge computing, green communication systems, wireless communication networks, and Internet-of-Things (IoT).
%
\end{IEEEbiography}


\vskip 0pt plus -1fil

\begin{IEEEbiography}{Mohammed Hirzallah}(S'15) received  the  B.Sc.
degree  in  electrical  engineering  from  the  Univer-
sity  of  Jordan,  Amman,  Jordan,  in  2011,  and  the
M.Sc. degree in electrical and computer engineering
from The University of Arizona, Tucson, AZ, USA,
in  2015,  where  he  is  currently  pursuing  the  Ph.D.
degree.  His  research  interests   include  the  design
and   analysis   of   wireless   protocols   and   systems.
He investigates  the problem of coexistence  between
heterogeneous  wire
less  networks.
\end{IEEEbiography}
\vskip 0pt plus -1fil

\begin{IEEEbiography}{Marwan Krunz}(S'93-M'95-SM'04-F'10) is the Kenneth VonBehren Endowed Professor in the Department of Electrical and Computer Engineering at the University of Arizona (UA). He also holds a joint (courtesy) appointment as a Professor in the Department of Computer Science. From 2008 to 2013, he was the UA site director of ``Connection One", an NSF Industry/University Cooperative Research Center (I/UCRC) that focused on RF and wireless communication systems and networks. During that period, the center included five participating sites (ASU, UA, OSU, RPI, and the University of Hawaii) and 26+ affiliates from industry and national research labs. Currently, Dr. Krunz co-directs the Broadband Wireless Access and Applications Center (BWAC), an NSF I/UCRC that includes UA (lead site), Virginia Tech, University of Notre Dame, University of Mississippi, Auburn University, and Catholic University of America. Along with NSF support, BWAC is currently funded by numerous companies and DoD labs, including Raytheon, Keysight Technologies, Alcatel-Lucent, Intel, L-3 Communications, Motorola Solutions, National Instruments, ONR, and others. BWAC aims at advancing the underlying technologies and providing cost-effective and practical solutions for next-generation (5G \& beyond) wireless systems, millimeter-wave communications, wireless cybersecurity, shared-spectrum access systems, full-duplex transmissions, massive MIMO techniques, and others. Dr. Krunz received the Ph.D. degree in electrical engineering from Michigan State University in July 1995. He joined the University of Arizona in January 1997, after a brief postdoctoral stint at the University of Maryland, College Park. In 2010, he was a Visiting Chair of Excellence (``Catedra de Excelencia") at the University of Carlos III de Madrid (Spain), and concurrently a visiting researcher at Institute IMDEA Networks. In summer 2011, he was a Fulbright Senior Specialist, visiting with the University of Jordan, King Abdullah II School of Information Technology. He previously held numerous other short-term research positions at the University Technology Sydney, Australia (2016), University of Paris V (2013), INRIA-Sophia Antipolis, France (2011, 2008, and 2003), University of Paris VI (LIP6 Group, 2006), HP Labs, Palo Alto (2003), and US West Advanced Technologies (1997).
\end{IEEEbiography}

\end{document}